\documentclass{emulateapj}

\def\gsim{\mathrel{\rlap{\lower 4pt \hbox{\hskip 1pt $\sim$}}\raise 1pt
\hbox {$>$}}}
\def\lsim{\mathrel{\rlap{\lower 4pt \hbox{\hskip 1pt $\sim$}}\raise 1pt
\hbox {$<$}}}

\begin{document}

\title{Signatures of A Companion Star in Type Ia Supernovae}

\author{
Keiichi Maeda\altaffilmark{1,2}, Masamichi Kutsuna\altaffilmark{3}, Toshikazu Shigeyama\altaffilmark{3}
}

\altaffiltext{1}{Department of Astronomy, Kyoto University, Kitashirakawa-Oiwake-cho, Sakyo-ku, Kyoto 606-8502, Japan; 
keiichi.maeda@kusastro.kyoto-u.ac.jp .}
\altaffiltext{2}{Kavli Institute for the Physics and Mathematics of the 
Universe (WPI), Todai Institutes for Advanced Study, University of Tokyo, 
5-1-5 Kashiwanoha, Kashiwa, Chiba 277-8583, Japan}
\altaffiltext{3}{Research Center for the Early Universe, School of Science, University of Tokyo, 7-3-1 Hongo, Bunkyo-ku, Tokyo 113-0033, Japan}

\begin{abstract}
While type Ia Supernovae (SNe Ia) have been used as precise cosmological distance indicators, their progenitor systems remain unresolved. One of the key questions is if there is a non-degenerate companion star at the time of a thermonuclear explosion of a white dwarf (WD). In this paper, we investigate if an interaction between the SN ejecta and the companion star may result in observable footprints around the maximum brightness and thereafter, by performing multi-dimensional radiation transfer simulations based on hydrodynamic simulations of the interaction. We find that such systems result in variations in various observational characteristics due to different viewing directions, while the predicted behaviors (redder and fainter for the companion direction) are opposite to what were suggested by the previous study. The variations are generally modest and within observed scatters. However, the model predicts trends between some observables different from observationally derived, thus a large sample of SNe Ia with small calibration errors may be used to constrain the existence of such a companion star. The variations in different colors in optical band passes can be mimicked by external extinctions, thus such an effect could be a source of a scatter in the peak luminosity and derived distance. After the peak, hydrogen-rich materials expelled from the companion will manifest themselves in hydrogen lines. H$\alpha$ is however extremely difficult to identify. Alternatively, we find that P$_{\beta}$ in post-maximum near-infrared spectra can potentially provide powerful diagnostics.  
\end{abstract}


\section{Introduction}

Type Ia supernovae (SNe Ia) are mature standardized candles, and have been playing a key role in the observational cosmology \citep{riess1998,permutter1999}. The SN Ia cosmology relies on an empirically derived relation between the peak luminosity and light curve decline rate (e.g., $\Delta m_{15}$ defined as a magnitude decrease from the peak to 15 days after), the so-called Phillips relation \citep{phillips1999}. This is also complemented by further relations between the luminosity (or decline rate) and the intrinsic colors (most frequently $B-V$), as essential in calibrating the external extinction \citep[e.g., ][and references therein]{folatelli2010}. 

However, the progenitor and explosion of SNe Ia are not yet fully understood. There has been a long debate about the progenitor system(s). The proposed systems are largely divided into two categories. One is called the single degenerate (SD) scenario, where a white dwarf (WD) accretes materials from its binary non-degenerate companion star, either a red giant (RG) or a main sequence (MS), to increase its mass to the (nearly) Chandrasekhar limit and ignites carbon near the center \citep[e.g., ][]{whelan1973,nomoto1982,hachisu1999}.  The companion stars could be even of different types \citep[e.g., ][]{wang2009,wheeler2012,liu2013}, but in this paper we mainly focus on RG and MS cases. The other one involves a merger of two WDs, and is called the double degenerate (DD) scenario \citep[e.g., ][]{iben1984,webbink1984}. The merger of WDs may result in a prompt explosion \citep{pakmor2010}, or else it may resemble to the final stage of the SD scenario with a larger accretion rate: a WD with the (nearly) Chandrasekhar  mass may evolve hydrostatically toward the central ignition \citep{yoon2007}, or such a system may experience an accretion induced collapse rather than SNe Ia \citep{saio1985}.  Different scenarios will lead to different explosion mechanisms, and thus understanding the origin of the diversity and relations in SN Ia luminosity and other observational properties relies on understanding the progenitor and evolution scenarios. Different scenarios may well predict a different evolution of SN properties as a function of the redshift, thus this is also a critical question in the SN Ia cosmology. 

New development on this issue has been achieved in the last few years from observational viewpoints. So far the strongest constraints have been placed by direct searches for surviving non-degenerate companion stars. A supernova remnant (SNR) 0509-67.5 in the Large Magellanic Cloud (LMC) was found to have no point sources down to $M_{V} \sim 8.4$ in the central region, ruling out both a RG and MS surviving companion for this particular SN \citep{schaefer2012}. It has been reported that a giant companion star is absent in SNR 1006 \citep{gonzalez2012} and SN 2011fe in M101 \citep{li2011}. 

Somehow model dependent, but still useful constraints on a possible companion star have been inferred by properties of nearby SNe Ia as well. Thermal energy deposited by a collision between the SN ejecta and the companion is predicted to produce detectable blue emission in the pre-max rising phase, especially for the RG companion case \citep{kasen2010}, while such an effect has not been clearly seen in a large sample of SNe Ia \citep{hyden2010}. In addition to the sign in the light curve, this collision is suggested to leave its sign in the spectrum. The hydrogen-rich envelope of the companion star is expected to be stripped away by the collision, being embedded mostly in the innermost, low-velocity part of the SN ejecta \citep{marietta2000}. Such hydrogen-rich material is suggested to produce H$_{\alpha}$ emission in the late phase \citep{mattila2005}, while there has been no sign of H$_{\alpha}$ in nearby SNe Ia so far \citep{mattila2005,leonard2007,shappee2013,lundqvist2013}. 

On the other hand, there are also observational indications for the SD scenario for at least a part of SNe Ia. There is a strong candidate for a surviving G-type dwarf in Tycho's SNR \citep{ruiz2004,bedin2013} \citep[but see also][]{ihara2007,kerzendorf2009}. The discovery of strongly interacting SNe Ia (SNe Ia exploding within dense CSM) favors the SD scenario for these SNe, specifically systems in which the companion is still in the non-degenerate phase at the time of the explosion \citep{hamuy2003,aldering2006,dilday2012}. In particular, \citet{dilday2012} discovered evidence for the traces of nova explosions preceding the SN, which had been predicted by the SD scenario. From this we expect that there are also cases where a WD with a non-degenerate companion explodes but without showing strong CSM interaction signals \citep{hachisu2012}. In sum, the issue is still controversy, and further study is required. Especially, the arguments based on SN properties can be model dependent \citep[e.g., see ][for uncertainties of the collision-induced emission in the pre-maximum stage ]{kutsuna2013a,kutsuna2013b}, and thus different ideas based on different physical processes and different observational strategies are quite useful. In this respect, we investigate the issue of how the maximum and post-maximum phases are affected by the existence of a companion star -- while this is the most easily accessible observations, indeed the model predictions so far are mostly restricted to the early rising phase (a few days after the explosion) or the late-time nebular phase (about an year after the explosion). There has been virtually only one study on this issue by \citet{kasen2004}. They predicted that the maximum spectrum is generally not sensitive to a viewing angle, while the SN looks blue and peculiar (i.e., 1991T-like) when viewed from the direction of the hole created by the ejecta-companion interaction. There were however some limitations in their study: (1) It has not been clarified if the overall (or angle-averaged) properties are affected as compared to the non-interaction case, (2) the prediction was made only for the maximum spectra, thus there were no specific features predicted for the post-maximum spectra, spectral evolution, or the multi-band light curves, (3) finally, the model was based on a toy model, which might be missing some ingredients important in hydrodynamics. 

In this paper, we explore the maximum and post-maximum properties of SNe in the optical through near-infrared (NIR) wavelength ranges, as a result of interaction between ejecta and a non-degenerate companion star. In \S 2, we summarize our hydrodynamic models and methods for radiation transfer simulations. In \S 3, we discuss overall properties in spectra and multi-band light curves. In \S 4, we discuss details on individual spectral features, colors, and their mutual relations predicted by the simulations. In \S 5, we investigate a possibility to detect hydrogen lines as diagnostics of the non-degenerate companion star. The paper is closed in \S 6 with conclusions and discussion. We describe details of radiation transfer simulation methodology in Appendix. 

\section{Method and Models}

\subsection{Hydrodynamic Models}

\begin{deluxetable}{cccc}
 \tabletypesize{\scriptsize}
 \tablecaption{Models
 \label{tab1}}
 \tablewidth{0pt}
 \tablehead{
   \colhead{Model}
 & \colhead{$M_2 (M_{\odot})$\tablenotemark{a}}
 & \colhead{$R_2 (10^{13} \ {\rm cm})$\tablenotemark{b}}
 & \colhead{$A (10^{13} \ {\rm cm})$\tablenotemark{c}}
}
\startdata
MS  & 1 & 0.01 & 0.03\\
RGa & 1 & 0.7  & 2\\
RGb & 1 & 1    & 3\\
\enddata
\tablenotetext{a}{The mass of the companion star.}
\tablenotetext{b}{The radius of the companion star.}
\tablenotetext{c}{The binary separation.}
\end{deluxetable}

\begin{figure*}
\begin{center}
        \begin{minipage}[]{0.95\textwidth}
                \epsscale{1.0}
                \plotone{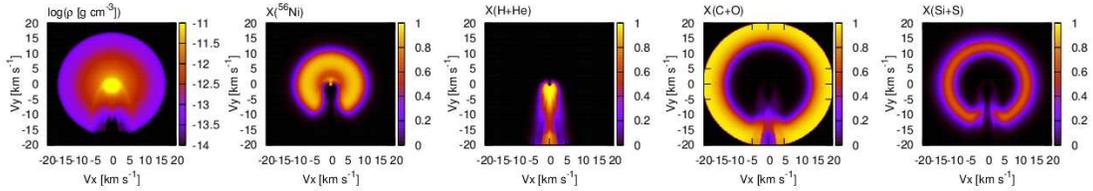}
        \end{minipage}
\end{center}
\caption
{The ejecta structure of Model RGa. The density is scaled to the value at 10 days since the explosion (when the ejecta are already in a homologous expansion). The companion was initially on the $-y$ direction (i.e., toward the bottom). The viewing angle $\theta$ is defined to be $\theta = 0$ in the companion direction ($-y$ in this figure), while $\theta = \pi$ in the opposite direction ($+y$). 
\label{fig1}}
\end{figure*}

Our input models for the radiation transfer simulations are taken from \citet{kutsuna2013a} \citep[see also ][]{kutsuna2013b}. In this section, we summarize main features of the models, and we refer \citet{kutsuna2013a} for further details of the models. 

These models are results of radiation hydrodynamic simulations of the collision between the expanding SN ejecta and a non-degenerate companion star. Thus, the initial configuration is specified by a few binary parameters, namely the type of the companion star and the separation between the WD and the companion. We take three models from their simulations, named Models MS, RGa, and RGb. Basic features of these models are summarized in Table 1. Model RGa represents a close binary system with a RG companion. The RG has $0.4 M_{\odot}$ of the He core and $0.6 M_{\odot}$ of the convective H-rich envelope. The separation is set to be $2 \times 10^{13}$ cm. Model RGb is the same with RGa, except for the separation being $3 \times 10^{13}$ cm. Model MS represents a close binary system with a MS companion. The companion MS mass is $1 M_{\odot}$ and the separation is $3 \times 10^{11}$ cm. The composition in the H envelope is set as follows: 75\% in H and 25\% in He. We ignore metal content in the hydrogen envelope (see \S 6). The radius of the companion is determined by a requirement that the companion star filled the Roche lobe just before the explosion.

Figure 1 shows the ejecta structure of Model RGa in the homologous expansion phase. The other models in the same phase are qualitatively similar to Model RGa, with only slight differences in details (e.g., in the opening angle of the hole created by the interaction). For details of the hydrodynamic behaviors, see \citet{kutsuna2013a}. The H-rich envelope mass stripped by the interaction is $\sim 0.4 M_{\odot}$ in all the three models. We note here that while the hydrodynamic behaviors are generally consistent with previous studies \citep[e.g., ][]{marietta2000}, the companion star in Model MS suffers from a large amount of hydrogen stripping. This is likely due to an insufficient computational resolution \citep{pakmor2008}, thus the amount of hydrogen in Model MS should be regarded as being overestimated \citep{kutsuna2013a,kutsuna2013b}. We note however that how much hydrogen is stripped away in the WD-MS system is still under debate \citep{liu2012}.

The W7 model \citep{nomoto1984} was used for the SN ejecta model. Accidentally, in \citet{kutsuna2013a} \cite[see also ][]{kutsuna2013b} the innermost stable Ni was counted as radioactive $^{56}$Ni in preparing the input model, thus this model has $\sim 0.8 M_{\odot}$ of $^{56}$Ni, which is larger than in the original W7 model. After mapping onto our numerical grids for the radiation transfer, the amount of $^{56}$Ni is $0.81M_{\odot}$ in our ejecta model while $0.59 M_{\odot}$ in the original W7 model. In any case, since the exact explosion mechanism is not yet clarified and $M$($^{56}$Ni) is within the observationally derived range of Branch-normal SNe Ia \citep{branch2006}, we take this `modified' W7 model as our reference model. Note that we are mainly aiming at investigating differences between the SN Ia with and without a non-degenerate companion, variations from different viewing directions -- thus details of the `reference' model are not important.  

In \citet{kutsuna2013a}, the collision between the SN ejecta and the companion star was simulated by two-dimensional radiation hydrodynamic simulations, with a few simplifications in the radiation transfer scheme (e.g., gray transfer, ignoring the bound-bound transition, a simplified $\gamma$-ray deposition scheme, flux-limited diffusion approximation). After the collision, the expanding ejecta (affected by the impact to the companion) reach the homologous expansion quickly, and the density structure in the homologous phase is little affected by radiation transfer effect. Thus we adopt the density and composition structures at 35 days after the explosion as our reference, and set the ejecta structure according to the homologous expansion to the initial time for the detailed radiation transfer simulations (typically 10 days after the explosion). While the impact dissipates the kinetic energy \citep{kasen2010,kutsuna2013a}, the resultant thermal energy is lost in a few days mostly due to the adiabatic loss. As such, in a few days since the explosion, the thermal condition is determined by the radioactive input. Thus, we neglect the thermal energy content due to the impact in our radiation simulations. 

In addition to these `companion-interaction' models, we also perform the radiation transfer simulations for the original W7 model {\em and} our own 1D reference model without interaction. The reference 1D model is constructed as follows - we extract the radial information from Model MS in the direction opposite to the companion, and this radial structure is mapped into all the directions in 3D space. This model represents the SN ejecta model without the interaction. This model is used for a fair comparison to the interaction models, better than the original W7 model. Reasons for this are (1) the SN ejecta models used for the interaction simulations have the distribution and mass of $^{56}$Ni slightly different from the original W7 model (see above), and (2) this is computed through the same hydrodynamic code with the interaction models, thus a possible numerical diffusion is taken into account in this reference model in the same manner as in the interaction models.

\subsection{Radiation Transfer}

We have performed radiation transfer simulations for the input models described in \S 2.1, mapped onto the 2D axi-symmetric coordinates with grid points $(n_r, n_\theta) = (50, 50)$, where $r$ and $\theta$ represent radial and polar angle coordinates. This spatial resolution is enough to resolve the major features in the ejecta structure arising from the interaction (Fig. 1), and also sufficient to resolve the spectral features arising from the photon Doppler shift (i.e., the radial spatial resolution corresponds to the photon Doppler shift of $\sim 300$ km s$^{-1}$ finer than typical spectral resolution in observations). We have used a multi-dimensional/frequency/epoch radiation transfer code developed by ourselves, {\em HEIMDALL (Handling Emission In Multi-Dimension for spectrAL and Light curve calculations)}. The full details of the code are presented in Appendix A, and in this section we will provide a summary of the simulation method and description specific to simulations in this paper. 

The code largely adopts prescriptions presented by \citet{lucy2005}, \citet{kasen2006}, and \citet{kromer2009}. The code solves radiation transfer for density and composition (taking into account radioactive decays) structures as a function of time given as an input model [$\rho (\vec r, t), X_i (\vec r, t)$]. The radiation field is solved with the Monte-Carlo (MC) method, where the radiation field is discretized into photon packets and the interactions between the radiation and matter are treated as individual microscopic events in the comoving frame \citep[][and references therein]{lucy2005}. To solve the radiation transfer, a mixed-frame approach is adopted, where the transformation from the comoving to the rest frames, and vice versa, automatically takes into account Doppler shift of radiation with respect to the matter. This is essential in the SN radiation transfer, since the large velocity gradient results in the wavelength shift of photons in the comoving frame even without interaction, and this wavelength shift can be much larger than the typical separations of the bound-bound transitions in the frequency space. Temperature at each position and time [$T (\vec r, t)$] is iteratively solved with radiation field (both in optical-NIR and $\gamma$-rays) within a time step, under the assumption of radiative equilibrium. Ionization and level populations are computed under the assumption of Local Thermodynamic Equilibrium (LTE). The new temperature is then used for an initial guess of the temperature in the next time step, and the radiation field at the end of a given time step is used as the initial radiation field in the beginning of the next time step. 

At the beginning of the simulation, MC packets representing $\gamma-$rays are created. They are assigned with the frequency, energy, spatial position, and emission epoch, following the radioactive decay chain of $^{56}$Ni $\to$ $^{56}$Co $\to$ $^{56}$Fe. The transfer of $\gamma$-rays are then followed with prescriptions given by \citet{maeda2006a}. The interactions include Compton scattering, photoelectric absorption, and pair creation. As a result, the energy deposition rate by $\gamma$-rays is obtained as a function of position and time. Together with the positron energy input, which is assumed to take place {\em in situ}, the heating/creation rate of `optical' (or thermal) photons is obtained. 

Using the energy deposition rate obtained through the $\gamma$-ray transfer, thermal photon packets are created. These are then followed by the MC simulation as described above. For the opacity to thermal (UV through NIR) photons, we adopt a standard set of the opacities largely used in the radiation transfer simulations in the expanding SN ejecta -- electron scattering, free-free, bound-free, and bound-bound transitions. In this paper, we adopt the expansion opacity prescription and two-level approximation for the discrete transitions. This introduces one parameter in dealing with the discrete transitions called thermalization parameter $\epsilon$. We adopt $\epsilon = 0.3$ in our simulations (see Appendix A). 

Throughout the simulation, frequency, energy, and directions of photon packets are recorded for those escaping the simulation region (i.e., SN ejecta) within the simulated time interval, together with the time of the escape. This information is summed up at the end of the simulations to reconstruct the escaping radiative flux from the system. In this process, the photons are binned into 100 polar directions from 0 to 180 degree spaced equally in the solid angle, and wavelength bins which are discretized into 3000 bins logarithmically spaced in frequency space from 100\AA\ to 20,000\AA. 
The time bins are divided into 30 from 10 days to 80 days, equally spaced in a logarithmic scale. Practically, we have performed the `first' MC transfer simulations with a smaller number of photon packets ($\sim 10^8$ in total) to converge  thermal conditions in ejecta. Then adopting this thermal structure, we have performed the `final' MC transfer with a larger number of photon packets ($\sim 10^9$) to obtain a sufficiently high Signal-to-Noise ratio in the resulting angle and time-dependent spectra. This large number of photon packets allows us to extract smooth spectra with high S/N - namely, on average the number of photons in each time-wavelength-angle bin is $\gsim 100$.

\section{Light Curves and Spectra}

\begin{figure*}
\begin{center}
        \begin{minipage}[]{0.95\textwidth}
                \epsscale{0.9}
                \plotone{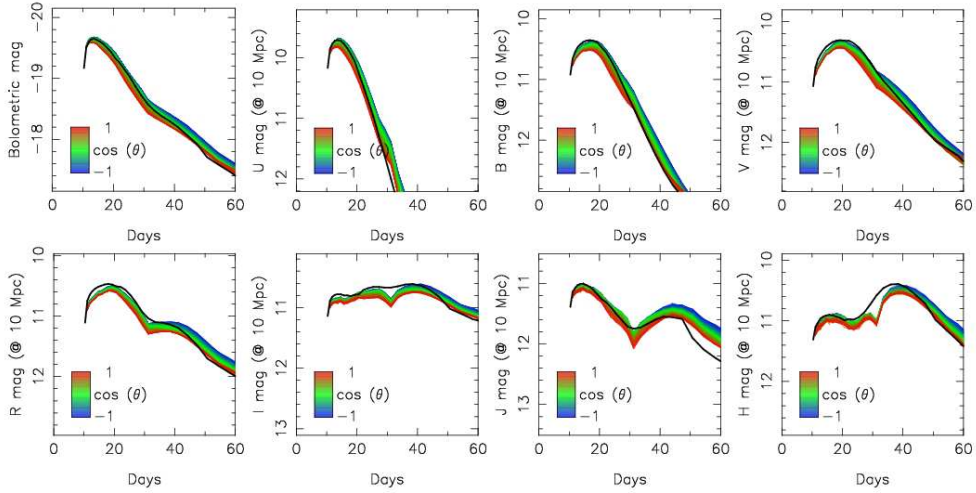}
        \end{minipage}
\end{center}
\caption
{Simulated multi-band light curves for Model RGa. The color coordinates indicate the light curves from different viewing directions (red for $\theta = 0$ and blue for $\theta = \pi$). The reference model curve is shown by the black-solid curve. The variation due to different viewing directions is modest, at 0.1 magnitude level in all the bands. Also, it is fainter if viewed from the companion direction ($\theta = 0$). 
\label{fig2}}
\end{figure*}

\begin{figure*}
\begin{center}
        \begin{minipage}[]{0.95\textwidth}
                \epsscale{0.9}
                \plotone{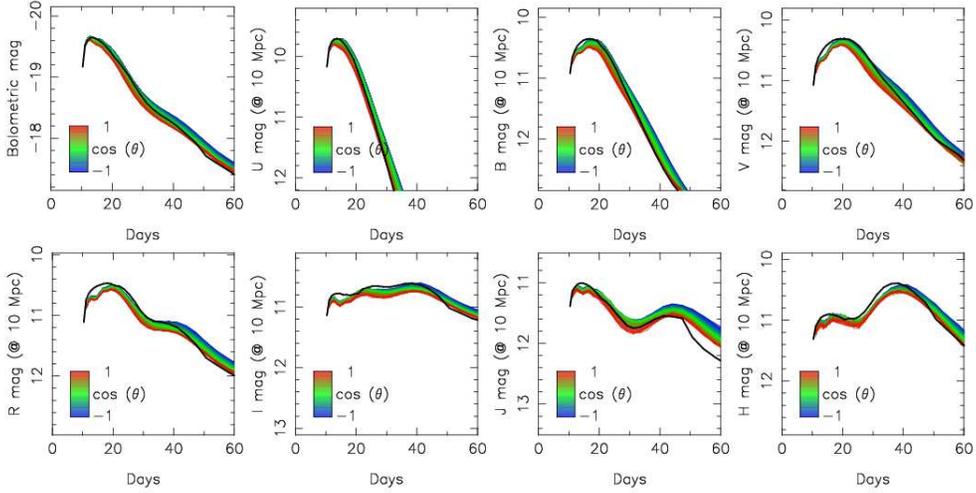}
        \end{minipage}
\end{center}
\caption
{Simulated multi-band light curves for Model RGb.
\label{fig3}}
\end{figure*}

\begin{figure*}
\begin{center}
        \begin{minipage}[]{0.95\textwidth}
                \epsscale{0.9}
                \plotone{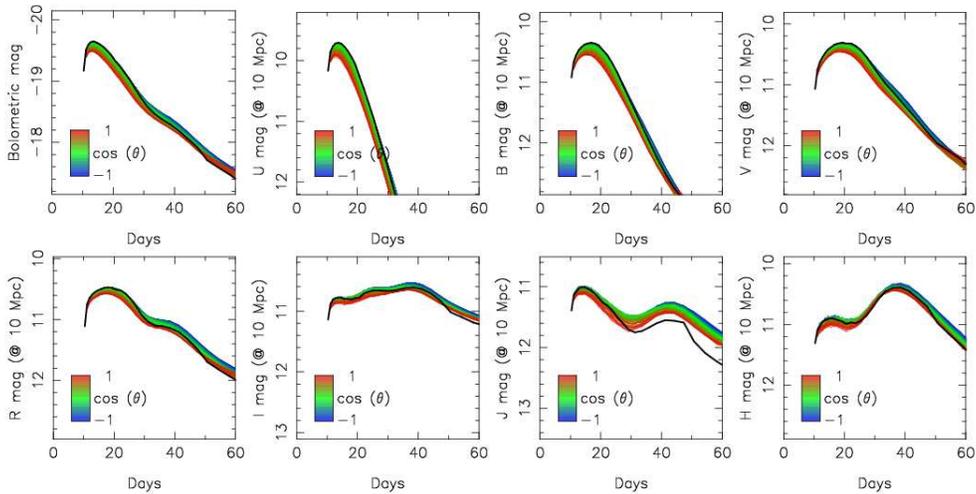}
        \end{minipage}
\end{center}
\caption
{Simulated multi-band light curves for Model MS.
\label{fig4}}
\end{figure*}

Hereafter, we present results of our simulations. We frequently comment on the viewing angle to an observer. In the following sections, we denote the viewing angle by $\theta$. The viewing angle $\theta$ is defined to be a polar angle as measured from the direction toward the companions star (or the hole). Namely, $\theta=0$ for an observer viewing the SN from the companion direction while $\theta = \pi$ for one in the opposite direction.

Figures 2--4 show synthetic multi-band light curves of Models RGa, RGb, MS, respectively. 
Those for our reference non-interaction model are also shown for comparison. Despite the large asymmetry in the ejecta structure (Fig. 1), the light curves of the interaction models are found to be very similar to those without interaction. Difference between different companion types is even smaller and there are virtually no difference visible down to the resolution in our simulations. We note that the deviation of the non-interaction model light curve in the $J$-band from the interaction models is presumably a numerical artifact, as well as a spiky feature in the light curves of Model RGa around 30 days. We note that this spiky feature (apparently non-converged temperature at this epoch) would not affect the later evolution, since we solve the temperature convergence at every time step.

\begin{figure*}
\begin{center}
        \begin{minipage}[]{0.95\textwidth}
                \epsscale{1.0}
                \plotone{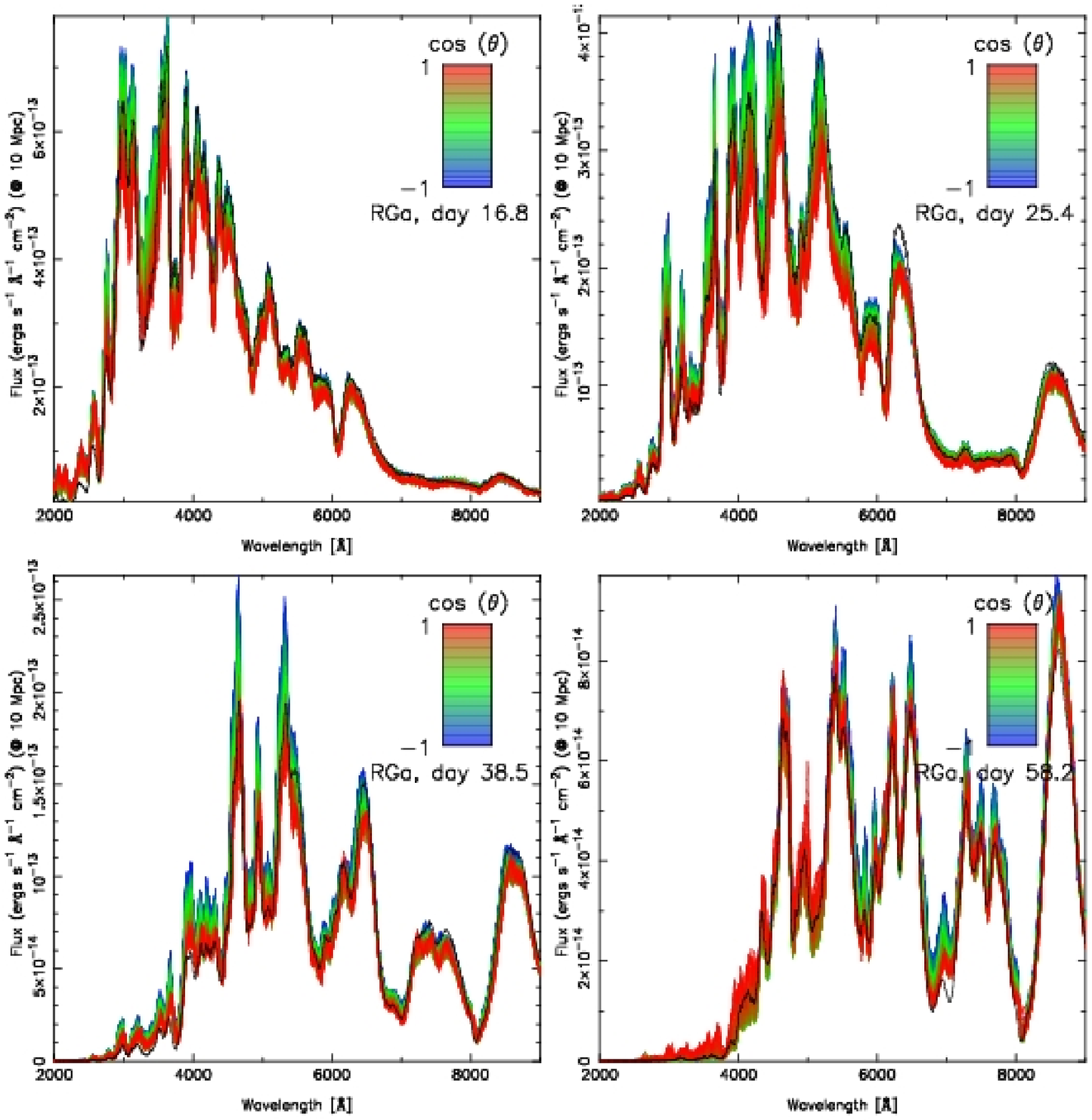}
        \end{minipage}
\end{center}
\caption
{Simulated spectra in optical wavelengths for Model RGa. The color coordinates indicate the spectra from different viewing directions (red for $\theta = 0$ and blue for $\theta = \pi$). When viewed from the companion (`hole') direction ($\theta=0$), the spectra are redder, with the smaller flux especially in the blue, than viewed from the opposite-side observer ($\theta=\pi$). The difference is at a moderate level so that the spectra would not be classified as peculiar. 
\label{fig5}}
\end{figure*}

A close inspection shows that around the peak the non-interaction model is almost identical to the interaction model viewed at $\theta \sim \pi$ (opposite to a companion or a hole). In the later phase, the non-interaction model is fainter than the angle-averaged mean light curves from the interaction model. This is consistent with the expectation -- early on the non-interacting side should look like a totally non-interaction model as one would not see the other side. Later on as the photon diffuses out one will eventually see effects of the interaction, and the bolometric luminosity should eventually follow the $\gamma$-ray deposition rate as the ejecta become transparent to optical photons. At this moment, the bolometric luminosity is approximated by a simple gamma-ray deposition rate \citep[see, e.g.,][]{maeda2003}, which is given as $L \propto M_{\rm ej}^2/E_{K}$ (where $L$ is the bolometric luminosity, $M_{\rm ej}$ and $E_{K}$ are the ejecta mass and the kinetic energy, respectively). Here, while $E_{K}$ should be unaffected by the interaction as the total energy is conserved, the ejecta mass is larger for the interaction case than the non-interaction case because of the addition of the envelope mass stripped from the companion. Namely, there are a larger amount of materials in the interaction case to absorb more $\gamma$-rays than the original ejecta without interaction. Naively, one would expect that the $\gamma$-ray deposition efficiency would be larger in the interaction model ($M_{\rm ej} \sim 1.8 M_{\odot}$) than the non-interacting model ($\sim 1.4 M_{\odot}$) by $\sim 60$\% if the deposition rate is simply scaled by the ejecta mass. This would lead to the late-time bolometric luminosity of the interaction model larger than the original by $\sim 0.5$ mag. This is enough to explain the difference in the late-time luminosity at $\sim 0.1 - 0.2$ magnitude level in the two models. Note that the difference as large as 0.5 magnitude is the maximum difference we expect, as the asymmetry in the ejecta in the interaction model is expected to lead to more effectively escaping $\gamma$-rays than in a spherical model with the same ejecta mass. 

The variation arising from different viewing directions is at the level of 0.1 mag, and an SN looks generally fainter for directions closer to the companion/hole ($\theta \sim 0$) in all the bands. As we have found little difference for different models, in the rest of the paper we mainly focus on Model RGa.

\begin{figure*}
\begin{center}
        \begin{minipage}[]{0.95\textwidth}
                \epsscale{1.0}
                \plotone{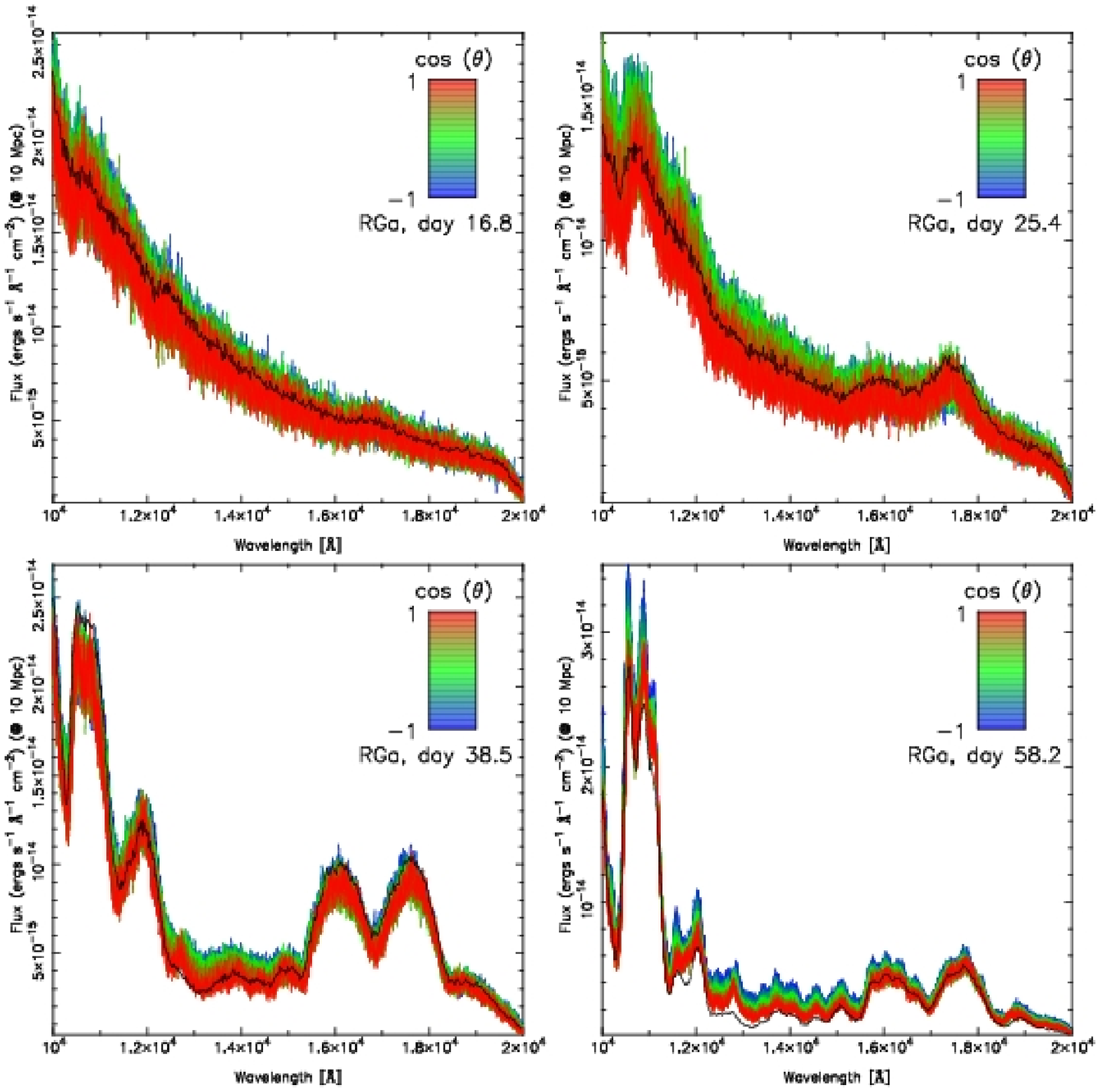}
        \end{minipage}
\end{center}
\caption
{Simulated spectra in NIR wavelengths for Model RGa. 
\label{fig6}}
\end{figure*}

Figures 5 and 6 show the synthetic spectral time sequence in the optical wavelength and in the NIR wavelength, respectively. As expected from the light curves, the interaction with the companion is found not to create any dramatic effects in the spectra as well. Namely, existence of a close binary non-degenerate companion does not leave detectable features in overall spectra around the maximum-light and thereafter (i.e., $10 - 80$ days since the explosion), and, in other words, this does not conflict with the observed uniformity of SN Ia spectra in these epochs.

Still, there is a difference. When viewed from the companion (`hole') direction  ($\theta=0$), the spectra are redder, with the smaller flux especially in the blue, than viewed from the opposite-side observer ($\theta=\pi$). The difference is at a moderate level so that the spectra would not be classified as peculiar and this SN would be classified into the same class irrespective of the viewing directions. Rather, this would create diversity, especially in the intrinsic color, within the same classification.

\begin{figure*}
\begin{center}
        \begin{minipage}[]{0.95\textwidth}
                \epsscale{1.0}
                \plotone{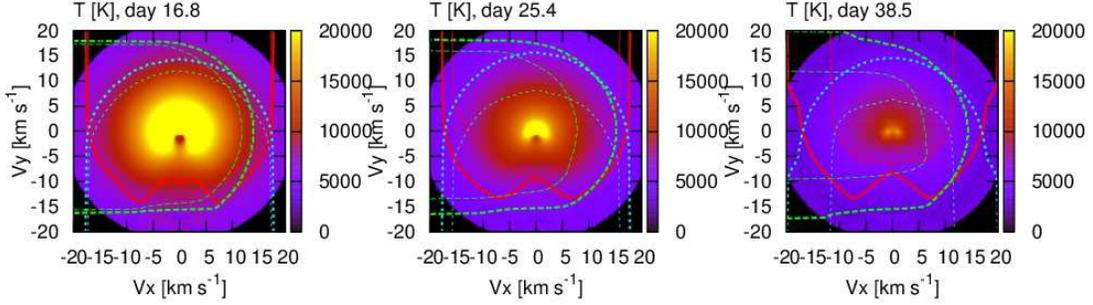}
        \end{minipage}
\end{center}
\caption
{Temperature distribution of model RGa. Also shown here are the U-band (thick) and R-band (thin) photosphere positions (as defined by $\tau = 2/3$). The photosphere is show for an observer at $\theta = 0$ (red), $\theta = \pi/2$ (green), and $\theta = \pi$ (blue). 
\label{fig7}}
\end{figure*}

\begin{figure*}
\begin{center}
        \begin{minipage}[]{0.95\textwidth}
                \epsscale{1.0}
                \plotone{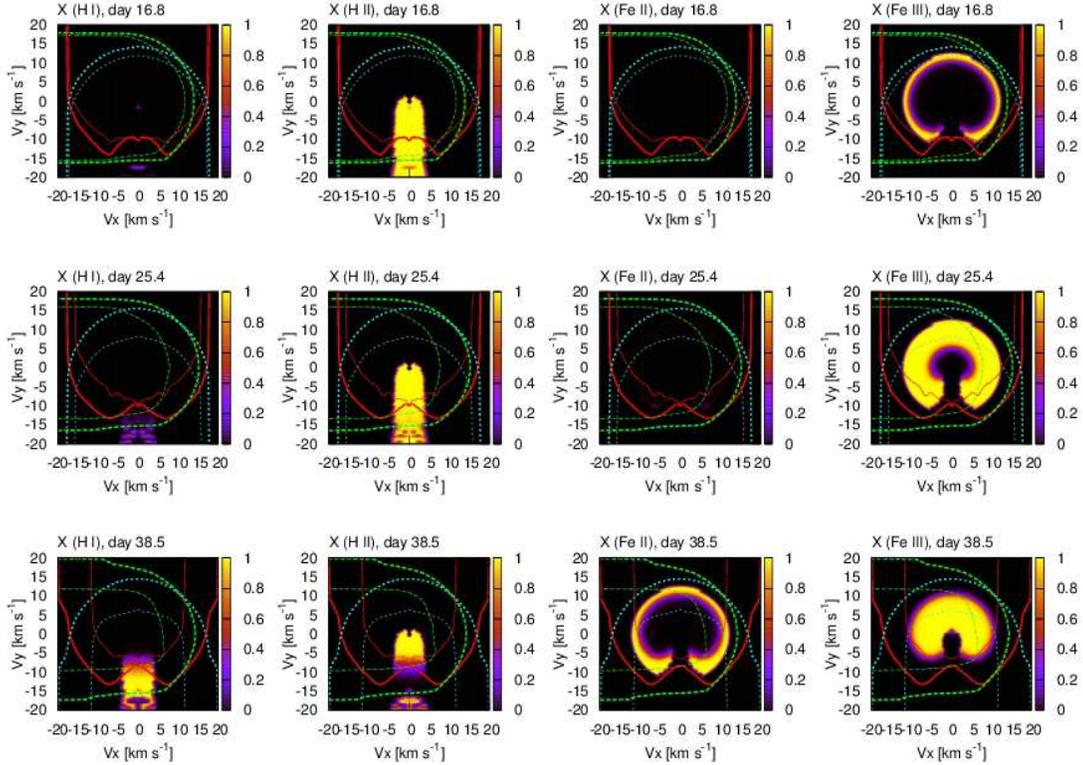}
        \end{minipage}
\end{center}
\caption
{Ionization Structure for Model RGa. Only the region where either hydrogen or the sum of Ni+Co+Fe exceeds 0.1 in mass fraction is shown. Also shown are the positions of the $U$-band (thick) and $R$-band photospheres (see the caption of Figure 7). 
\label{fig8}}
\end{figure*}

The temperature and ionization structures are shown in Figures 7 and 8, respectively. Overall, temperature is lower on the side of the companion, due to a smaller amount of the heating source ($^{56}$Ni) and also due to a smaller amount of absorbing matter (heavy elements, especially Fe-peak elements) on this side than the others. 

We note that our results are qualitatively different from those found by \citet{kasen2004}, who predicted that the SN is bluer and blighter (especially in the shorter wavelengths) when viewed at $\theta \sim 0$. The situation that an observer at $\theta \sim 0$ views directly at the $^{56}$Ni-rich region was a main cause of the predicted behavior by \citet{kasen2004}. \citet{kasen2004} performed a snap-shot spectral synthesis for a maximum-spectrum based on a toy model. In their model, they mimicked the outcome of the interaction as the SN ejecta with a hole represented by a (nearly) constant opening angle. We note that their model underestimates the amount of material in the `hole' -- a large amount of the SN ejecta, either C+O or Si-rich layer, fill up the hole left by the interaction, and also the H-rich companion materials are naturally filling the hole as well. Thus the `hole' is not really a vacuum. The existence of these materials does not allow the photosphere to quickly recede to the bottom of the hole (i.e., $^{56}$Ni-rich central region). Even with only the H-rich envelope from the companion intruding into this region, electron scattering can become significant (especially for the RG case) to clip the photosphere at a relatively high velocity -- with $\sim 0.5 M_{\odot}$ of fully ionized hydrogen materials (with 70\% in mass fraction) confined within a sphere below 1,000 km s$^{-1}$, the electron scattering optical depth is estimated to be $\tau \sim 2,700 \ (t/20 \ {\rm days})^{-2}$. Thus, this will hide $^{56}$Ni-rich region from the line-of-sight to an observer at $\theta \sim 0$ until the hydrogen recombines. The hydrogen is kept nearly fully ionized during the epochs of interest in this paper (Fig. 8) especially in the low velocity region, where $\gamma$-rays and optical photons are absorbed efficiently due to the large density. 

In our situation based on the hydrodynamic model, the temperature at the photosphere at $\theta \sim 0$ tends to be lower than that at $\theta \sim \pi$, since the companion direction ($\theta \sim 0$) is blocked by the $^{56}$Ni-free materials. This direction lacks Fe-peak elements, accordingly the photosphere at the $U$-band is at a low velocity. The position of the photospheres at the $R$-band is not extremely sensitive to the viewing direction, supporting the interpretation that the main difference in the photosphere in the $U$-band is caused by the different amount of Fe-peak elements that are sources of opacity especially in blue bands. The $U$-band photosphere does not quickly recede in the velocity space as compared to the $R$-band, for observers viewing from any directions, that reflects the increasing opacity in the blue at the later epochs due to recombination of Fe-peak elements (Fig. 8). 

\begin{figure}
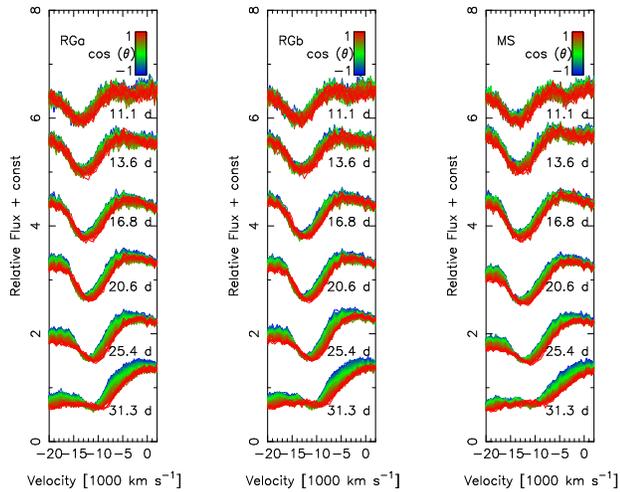

\begin{center}
        \begin{minipage}[]{0.15\textwidth}
                \epsscale{1.0}
                \plotone{f9a.eps}
        \end{minipage}
        \begin{minipage}[]{0.15\textwidth}
                \epsscale{1.0}
                \plotone{f9b.eps}
        \end{minipage}
        \begin{minipage}[]{0.15\textwidth}
                \epsscale{1.0}
                \plotone{f9c.eps}
        \end{minipage}
\end{center}
\caption
{Si II 6355 in the simulated spectra. The color coordinates indicate the spectra as viewed from different viewing directions (red for $\theta = 0$ and blue for $\theta = \pi$). Initially at $\sim 10$ days the Si II profile is similar for observers at any directions, then observers at smaller $\theta$ (toward the companion direction) will observe the Si II at progressively lower velocity than in the opposite direction. 
\label{fig9}}
\end{figure}

Figure 9 shows evolution of the spectral region around Si II 6355. Around the peak luminosity (i.e., $\sim 15$ days since the explosion), the absorption minimum (and the emission peak) is at longer wavelength (i.e., lower velocity) for observers at smaller $\theta$, as is consistent with the result by \citet{kasen2004}. We find that the predicted temporal evolution is also different for different viewing direction \citep[see also][]{kutsuna2013a,kutsuna2013b}. Indeed, at $\sim 10$ days after the explosion (i.e., about a week before the $B$-band maximum), the line profiles are not sensitively dependent on the viewing direction. After that, the line velocity decreases more quickly for an observer at $\theta \sim 0$, thus leading to progressively lower velocity for this direction. This temporal behavior is also different from that predicted by \citet{kutsuna2013a}, who predicted the lower velocity of the Si II for $\theta \sim 0$ than for $\theta \sim \pi$ already well before the maximum. Figure 10 demonstrates how the line profiles are different for different viewing directions at 16.8 and 38.5 days after the explosion. At 38.5 days, it is not easy to identify the Si II in the spectra. We note that the line profile at $\sim 6,150 - 6,200$\AA\ is different for different viewing directions, showing a small peak for $\theta \sim 0$ but not for $\theta \sim \pi$. This wavelength is influenced by the material moving at the velocity of $\sim 7,000 - 10,000$ km s$^{-1}$. Figure 10 (right panel) shows that Si II 6355 affects this wavelength range differently for different viewing directions. 

Figure 11 shows the Sobolev optical depth of Si II together with the $R$-band photosphere. Around the peak luminosity, the peak in the opacity distribution is on average at lower line-of-sight velocity for $\theta=0$, because the highest velocity materials are missing in this direction. At the epoch of the peak luminosity (at day 16.8), this particular optical depth is larger at $\theta=0$, as the lower temperature there prefers Si II over Si III. As time goes by, the temperature decrease induces the recombination of Si II to Si I. The ejecta on the companion side ($\theta \sim 0$) does not keep the high optical depth of Si II at day 38.5 while the ejecta on the opposite side ($\theta \sim \pi$) still maintains sufficient amount of Si II to keep its line velocity as high as $\sim 8,000$ km s$^{-1}$. As a result, at this epoch the flux around 6,200\AA\ is suppressed for $\theta \sim \pi$ but not for $\theta \sim 0$. In principle, this small but varying feature in Si II 6355 could be a powerful signature of the presence of a companion star, and this is further discussed in \S 6.

\begin{figure}
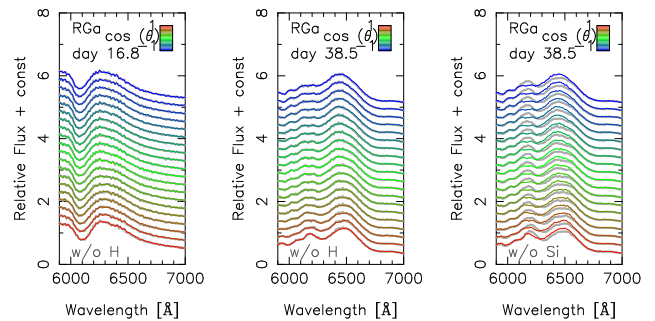

\begin{center}
        \begin{minipage}[]{0.15\textwidth}
                \epsscale{1.8}
                \plotone{f10a.eps}
        \end{minipage}
        \begin{minipage}[]{0.15\textwidth}
                \epsscale{1.8}
                \plotone{f10b.eps}
        \end{minipage}
        \begin{minipage}[]{0.15\textwidth}
                \epsscale{1.8}
                \plotone{f10c.eps}
        \end{minipage}
\end{center}
\vspace{2cm}
\caption
{The simulated spectra around Si II 6355. The color coordinates indicate the prediction for different viewing directions (red for $\theta = 0$ and blue for $\theta = \pi$). In this figure, the spectra for observers at different directions are added with an additional offset ($\theta = 0$ to $\pi$, from bottom to top). In he left two panels, the color curves are the model curves with hydrogen, while the gray curves are without hydrogen. The color curves shown in the right panel is identical to the one in the middle (i.e., model RGa at day 38.5 since the explosion) while the gray curves are without silicon. H$_{\alpha}$ is observationally not detectable in both epochs. The viewing angle dependence is small around the $B$-band maximum, while the variation arising from Si II 6355 becomes visible in the later epoch. 
\label{fig10}}
\end{figure}

\begin{figure*}
\begin{center}
        \begin{minipage}[]{0.95\textwidth}
                \epsscale{1.0}
                \plotone{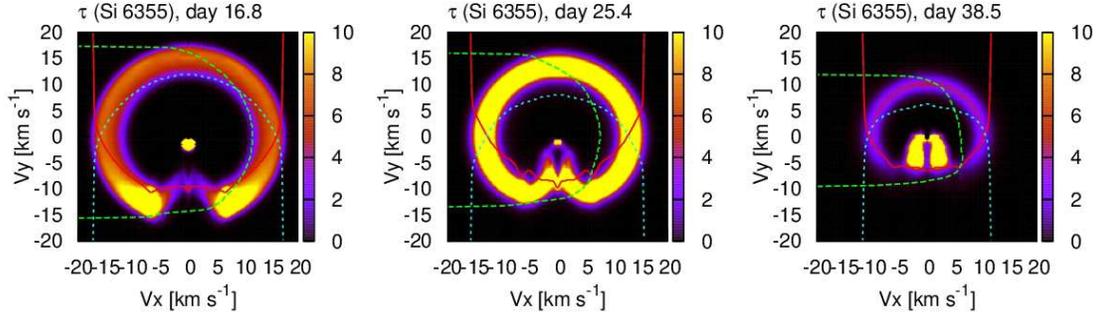}
        \end{minipage}
\end{center}
\caption
{Sobolev optical depth of Si II 6355 for Model RGa. Shown here is the R-band photosphere ($\tau = 2/3$) for an observer at $\theta = 0$ (red), $\theta = \pi/2$ (green), and $\theta = \pi$ (blue).
\label{fig11}}
\end{figure*}

\begin{figure*}
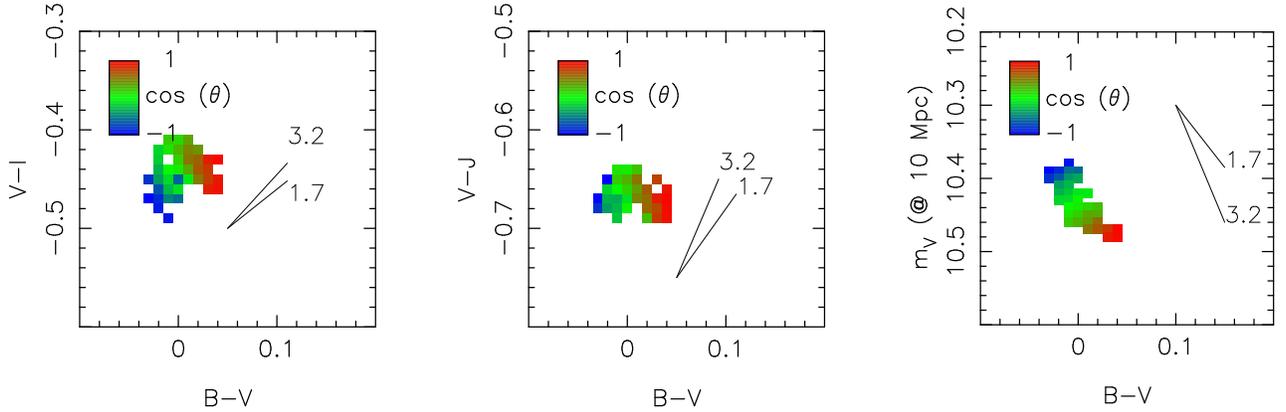

\begin{center}
        \begin{minipage}[]{0.32\textwidth}
                \epsscale{1.0}
                \plotone{f12a.eps}
        \end{minipage}
        \begin{minipage}[]{0.32\textwidth}
                \epsscale{1.0}
                \plotone{f12b.eps}
        \end{minipage}
        \begin{minipage}[]{0.32\textwidth}
                \epsscale{1.0}
                \plotone{f12c.eps}
        \end{minipage}
\end{center}
\caption
{The predicted relations in photometric properties for Model RGa. The color coordinates indicate the prediction for different viewing directions (red for $\theta = 0$ and blue for $\theta = \pi$). The external-extinction vector is shown for $R_V = 3.2$ and $1.7$. The variation in each quantity is at most at the level of 0.1 magnitude. The trend seen in the optical properties is similar to the effect of external extinction, while it is not the case in NIR. 
\label{fig12}}
\end{figure*}

\section{Spectral Features and Color}
In \S 3, we showed that the overall light curve and spectral behaviors are not much affected by the existence of a non-degenerate companion star. Still, we find a small variation for different viewing directions, and in this section we investigate more details on this issue. 

Figure 12 shows the variations in the colors and the $V$-band magnitude, and their relations to the $B-V$ color obtained for Model RGa. The variation in each quantity is at the level of 0.1 mag or even smaller, thus such an effect is difficult to see in current observations, and it is practically impossible to disentangle this effect from other possible sources of the scatter. Relations between different observables, however, have interesting implications. Within the optical range, the relations between different colors are indeed similar to the effect of external extinction. This is also true for relations between the color and absolute magnitude. This means that the viewing angle variation can mimic the external extinction, and could be a part of the origins of the scatter in the luminosity calibration \citep[see, e.g., ][]{maeda2011}. This degeneracy could be solved by adding the NIR information because the optical-NIR colors are found not to follow the extinction vector. 

\begin{figure}
\begin{center}
        \begin{minipage}[]{0.45\textwidth}
                \epsscale{0.9}
                \plotone{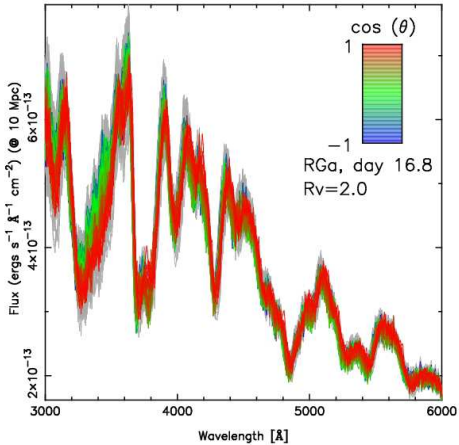}
        \end{minipage}
\end{center}
\caption
{A demonstration of how the intrinsic color difference of Model RGa as arising from different viewing directions can be mimicked by the external extinction. The original spectra for various viewing directions are shown in gray, while the ones corrected for the `hypothesized' extinction are shown in colors. 
\label{fig13}}
\end{figure}

The possible effect in the extinction estimate is demonstrated in Figure 13. Here, we show how the intrinsic color dispersion predicted for model RGa can be mimicked by the external extinction. We {\em hypothesized} here (while we know it is not the case) that the color variation here is entirely due to a different amount of the extinction, and convolve the external extinction associated with the model $B-V$ color assuming $R_V =2.0$ as is typically derived for SNe Ia. Figure 13 shows that, by doing this way the dispersion in the optical spectra is reduced -- namely, if one observes model RGa from various directions, one would associate the diversity in color to the external extinction. This would introduce a systematic error in the derived absolute magnitude at $0.1$ magnitude level. 

\begin{figure*}
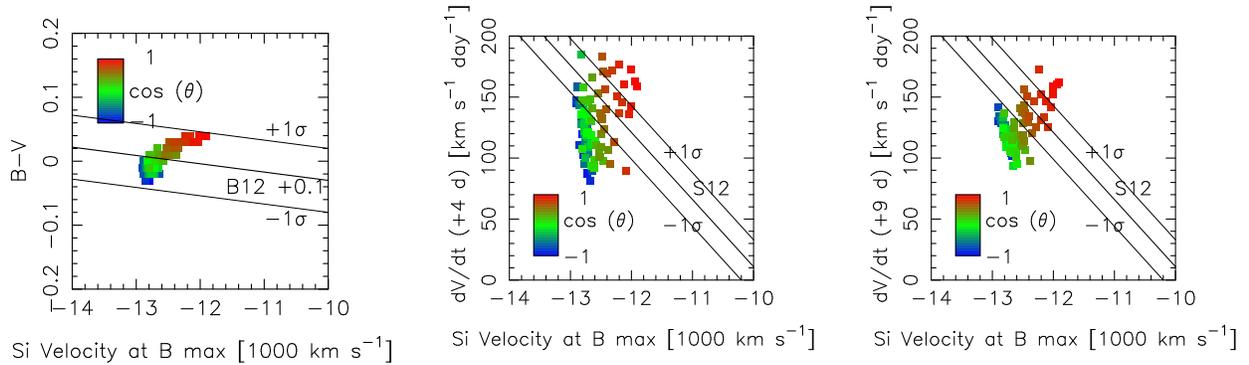

\begin{center}
        \begin{minipage}[]{0.3\textwidth}
                \epsscale{1.0}
                \plotone{f14a.eps}
        \end{minipage}
        \begin{minipage}[]{0.3\textwidth}
                \epsscale{1.0}
                \plotone{f14b.eps}
        \end{minipage}
        \begin{minipage}[]{0.3\textwidth}
                \epsscale{1.0}
                \plotone{f14c.eps}
        \end{minipage}
\end{center}
\caption
{The predicted relations in spectroscopic properties for Model RGa. The color coordinates indicate the prediction for different viewing directions (red for $\theta = 0$ and blue for $\theta = \pi$).  The solid lines are for the observationally derived relations with the $1\sigma$ statistical errors for the velocity--color \citep[with an additional offset; ][: B12]{blondin2012} and for the velocity--velocity gradient \citep[][: S12]{silverman2012}. The velocity gradient in the model is defined as the gradient in two epochs, where the first epoch is set at the $B$-band maximum and the second epoch is set either at 4 days or 9 days since the $B$-band maximum. The relations due to the viewing angle diversity do not follow the observed relations. The expected diversities are within the observed scatters at $1 - 2\sigma$ level. 
\label{fig14}}
\end{figure*}

One of the biggest effects of the interaction is that on the predicted velocity of absorption lines (\S 3). Indeed, the predicted relation between the velocity at the maximum-light and the $B-V$ color, and the one between the velocity and the velocity gradients are different than the observationally found relation. It has been reported that the absorption velocity around the maximum is correlated with $B-V$ in a way that the higher velocity SNe have redder intrinsic color \citep[][]{foley2011,maeda2011,blondin2012}. Figure 14 shows the predicted relation between the Si II velocity and $B-V$ color, overlapped with the observed trend from \citet{blondin2012} but added an offset of 0.1 mag. Note that observationally the zero-point in the intrinsic color is dependent on the determination of the reddening and also on the sample since the color is also dependent on $\Delta m_{15}$, and theoretically the absolute scale can be more sensitive to numerical details than the `relative' values (see Appendix B). Here we are interested in the trend and diversity. Figure 14 shows that the predicted trend is different from the observed trend, while the variation is still within the $1\sigma$ scatter of the observationally derived relation. 

There is also an observationally found relation between the velocity and `velocity gradient'. The velocity gradient is a measure of how quickly the absorption velocity decreases as a function of time \citep{benetti2004,benetti2005}. The velocity gradient is larger (i.e., the velocity decreases more quickly) for SNe with larger velocity at maximum \citep[e.g., ][]{silverman2012}. Figure 14 shows the relation predicted for Model RGa as compared to the observationally derived relation from \citet{silverman2012}. Observationally the velocity gradient is defined by a linear or higher order fit to a light curve just after the $B$-band maximum \citep{benetti2004,benetti2005,blondin2012}. The absolute values of the velocity gradient for specific objects are dependent on the time interval for the fit as well as the fitting function while overall tendency of larger gradient for higher-velocity SNe is not sensitive to these choices \citep[e.g., ][]{blondin2012}. Here, the velocity gradient in the model is defined as the gradient in two epochs for the sake of simplicity, where the first epoch is set at the $B$-band maximum. To check the robustness of the result, we vary the second epoch, and 
it is set either at 4 days or 9 days since the $B$-band maximum in Figure 14. While we are forced to adopt the second epoch not very late (as the model does not reproduce the 
observed Si II profile around day 25 and thereafter; see Appendix A), we see that our results are not affected much by the definition of the velocity gradient (see Figure 14). 

The velocity predicted for Model RGa is as high as those of the high velocity (or high velocity gradient) SNe, and the dispersion arising from different viewing angle alone does not explain the observed range of the Si II velocity (i.e., $10,000 - 12,000$ km s$^{-1}$ for the low velocity SNe, and reaching to $\sim 16,000$ km s$^{-1}$ for the high velocity SNe). Namely, the viewing angle effect arising from the asymmetry introduced by the ejecta-companion interaction is not a main cause of the diversity in the velocity \citep[see, e.g., ][for a possible origin of the diversity arising from the asymmetry in the explosion itself]{maeda2010a}. Moreover, the viewing angle effect here predicts the trend different than the observed one (Fig. 14). As such, the observed relation can, in principle, be used as a constraint on the existence of a non-degenerate companion star. The dispersion predicted for model RGa indeed exceeds the nominal $1\sigma$ error in the observed relation. There are, however, quite a number of outliers in this relation \citep[e.g., see Fig. 6 of ][]{silverman2012}, thus the present sample does not strongly reject the existence of a RG companion for a majority of SNe Ia. In any case, this relation is potentially a strong diagnostics to limit the fraction of SNe Ia with a non-degenerate companion star. 

In the above arguments, we have dealt with the small differences, due to various viewing directions, at 0.1 magnitude level. A question is if our simulations are accurate enough down to this level. In Appendix B, we discuss this numerical accuracy issue in details with test calculations, and conclude that the results are not numerical artifacts.

\section{Hydrogen Lines}

\begin{figure*}
\begin{center}
        \begin{minipage}[]{0.95\textwidth}
                \epsscale{0.7}
                \plotone{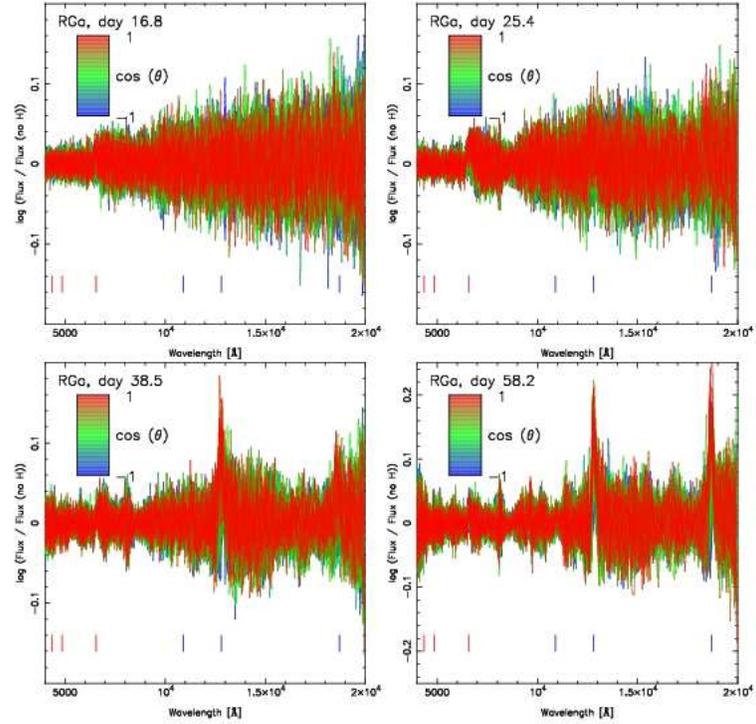}
        \end{minipage}
\end{center}
\caption
{The ratio of the spectral flux with and without hydrogen lines for Model RGa. The color coordinates indicate the models for different viewing directions (red for $\theta = 0$ and blue for $\theta = \pi$). The lines on the bottom show the rest wavelength positions of Balmer series ($\alpha$, $\beta$, $\gamma$) and Paschen series ($\alpha$, $\beta$, $\gamma$). In this plot, the simulated spectra are averaged in three time bins and three wavelength bins to increase the Signal-to-Noise ratio. In later phases, the variations in P$_{\alpha}$ and P$_{\beta}$ arising from different viewing directions clearly exceed the MC noise. 
\label{fig15}}
\end{figure*}

\begin{figure*}
\begin{center}
        \begin{minipage}[]{0.95\textwidth}
                \epsscale{0.7}
                \plotone{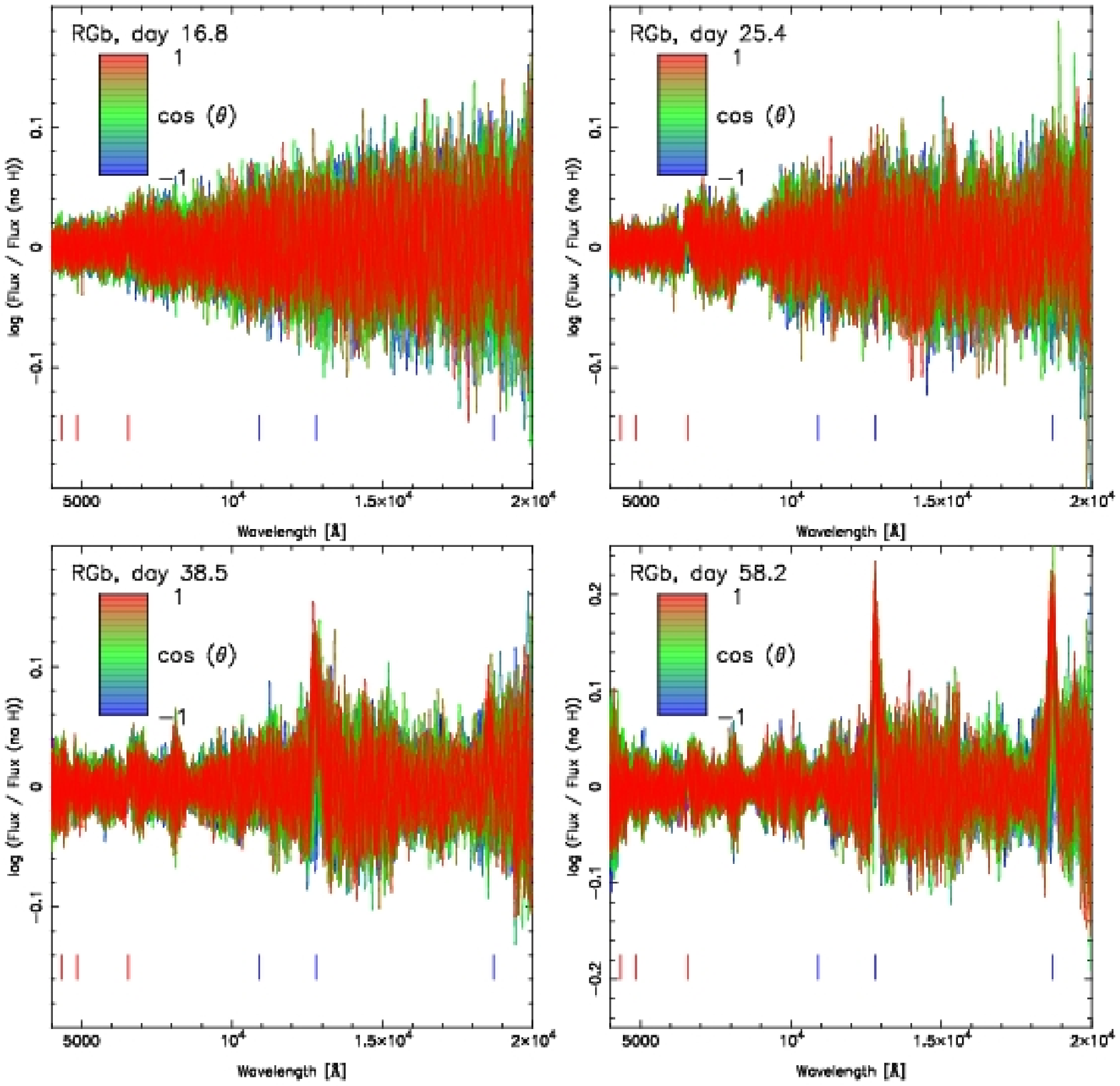}
        \end{minipage}
\end{center}
\caption
{The ratio of the spectral flux with and without hydrogen lines for Model RGb.
\label{fig16}}
\end{figure*}

\begin{figure*}
\begin{center}
        \begin{minipage}[]{0.95\textwidth}
                \epsscale{0.7}
                \plotone{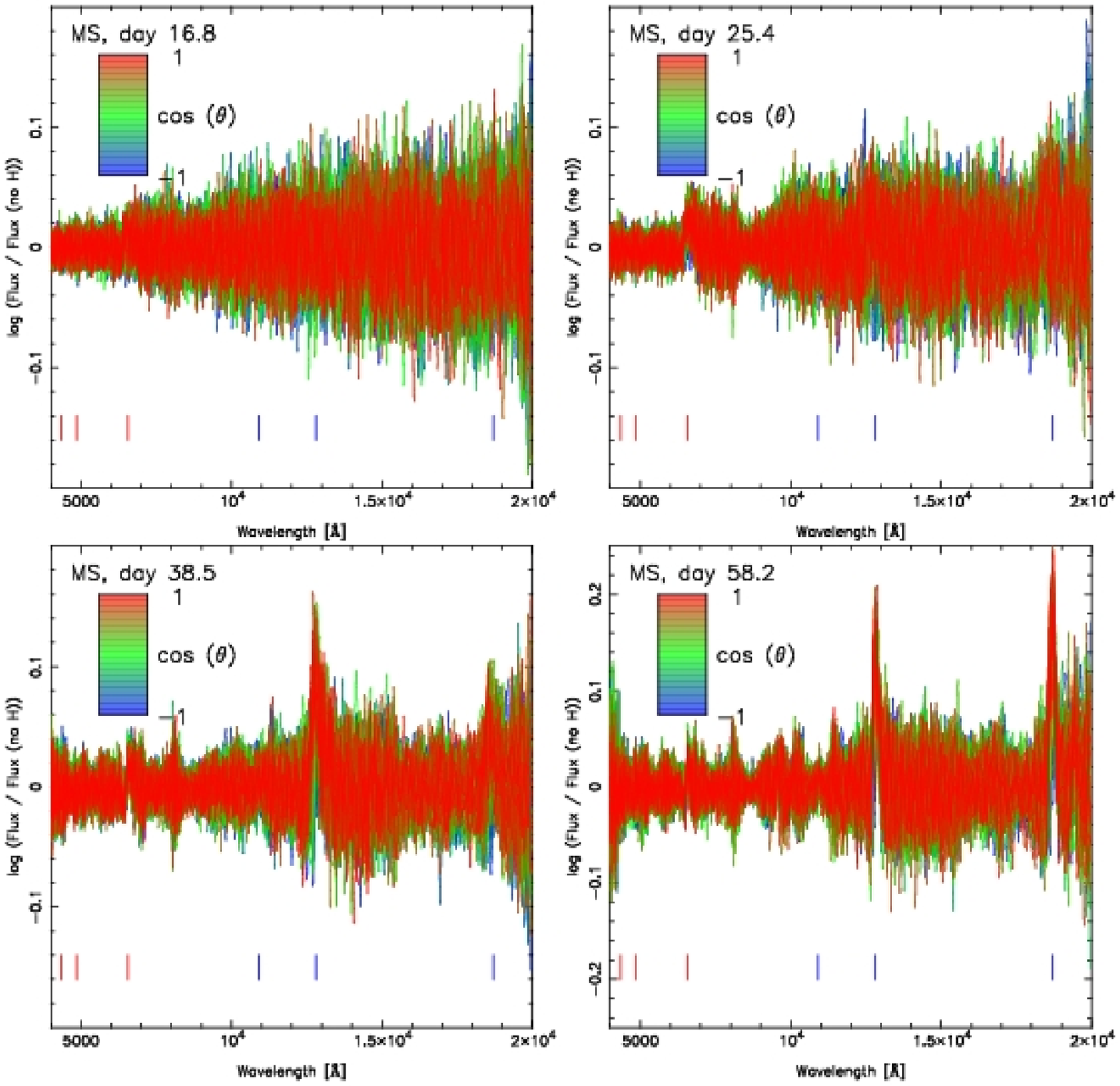}
        \end{minipage}
\end{center}
\caption
{The ratio of the spectral flux with and without hydrogen lines for Model MS.
\label{fig17}}
\end{figure*}

\begin{figure}
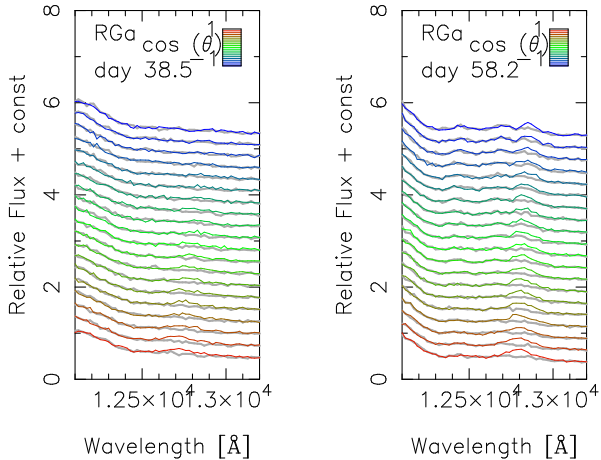

\begin{center}
        \begin{minipage}[]{0.23\textwidth}
                \epsscale{1.0}
                \plotone{f18a.eps}

        \end{minipage}
        \begin{minipage}[]{0.23\textwidth}
                \epsscale{1.0}
                \plotone{f18b.eps}
        \end{minipage}
\end{center}
\caption
{P$_{\beta}$ in the simulated spectra. In this figure, the spectra for an observer at different directions are added with an additional offset ($\theta = 0$ to $\theta = \pi$, from bottom to top). The color curves are the model curves with hydrogen, while the gray curves are without hydrogen. P$_{\beta}$ is present in these epochs, showing variations in the flux and profile for different viewing directions. 
\label{fig18}}
\end{figure}

The hydrogen-rich matter stripped off from the companion star, being embedded in the SN ejecta, has been regarded to be a key to probing (or rejecting) the existence of a non-degenerate companion star at the time of the SN Ia explosion. A large fraction of the stripped-off hydrogen is embedded near the center of the SN ejecta at low velocities which are only visible in the late phases when the SN ejecta become fully transparent. As such, searching for the H$_{\alpha}$ emission in late-time spectra has been suggested and performed for a few SNe Ia \citep{mattila2005,leonard2007,lundqvist2013,shappee2013}. A smaller amount of hydrogen are also distributed at high velocities \citep{marietta2000}, which could be in principle probed by more easily accessible maximum and post-maximum spectra. However, the latter issue has not been quantitatively investigated in the past. Lentz et al. (2002) investigated if the H-rich materials mixed into the high velocity part of the SN ejecta ($\gsim 15,000$ km s$^{-1}$) could be detected in the pre-maximum spectra,  through 1D radiation transfer simulations. They concluded that the signatures are expected to be stronger for the earlier phases, but the signals are generally weak. Recently, \citet{kutsuna2013a} suggested that one may see H$_{\alpha}$ even just after the maximum light (i.e., $\gsim 10$ days after the explosion) for the interaction with a non-degenerate companion, based on the hydrodynamic simulation and simplified radiation transfer. Here, we investigate this issue -- based on the same hydrodynamic simulation models with \citet{kutsuna2013a} and detailed radiation transfer, we investigate if the existence of hydrogen is visible in the maximum and post-maximum spectra (i.e., up to about a month since the $B$-band maximum). 

Figures 15--17 show the ratio of the synthetic spectra for the same model(s) but with and without hydrogen included in the radiation transfer. To create the hydrogen-free reference spectra, we performed the same radiation transfer simulation based on the temperature structure obtained through the original calculations, but setting the bound-bound and bound-free hydrogen opacities zero by hand. In Figures 15-17, the model results for different directions are shown by different colors (red for $\theta = 0$, and blue for $\theta = \pi$). Since our spectra are extracted from the MC-simulation by binning the emerging photon packets, we suffer from the MC noise. Especially, when the flux is smaller then the noise level becomes larger. For example, this is seen in the larger noise level at the longer wavelength especially apparent for the maximum spectra, or at the wavelength corresponding to the Ca II NIR absorption especially in the later-phase spectra. If there is a `real' feature produced by hydrogen above the level of the MC noise, the ratio at the corresponding wavelength should show imbalance with respect to unity (or zero in the logarithmic scale), namely the emission appears as the increase in the ratio (above unity) while the absorption appears as the decrement in the ratio (below unity). As such, a P-Cygni profile in the flux spectra should also appear in the same way in the `ratio spectra' shown in Figures 15--17.

H$_{\alpha}$ is seen in these figures (above the MC noise). Irrespective of the epoch, the ratio at the H$_{\alpha}$ is at 10\% level (but note that this is probably an overestimate for Model MS). Namely, if the observed Signal-to-Noise ratio exceeds this level and {\em if one knows the hydrogen-free reference spectrum a priori}, one may detect hydrogen through H$_{\alpha}$. We thus confirmed the suggestion by \citet{kutsuna2013a}, while the latter condition was not discussed by them. We investigate this issue later in this section.

We find from these figures that the hydrogen lines in NIR could provide potentially much stronger signals than H$_{\alpha}$. Around the $B$-band maximum brightness, strengths of P$_{\alpha}$ and P$_{\beta}$ are below the noise level of our MC simulation in the corresponding wavelength ($\sim$20\% level). As time goes by the signal becomes stronger and exceeds the MC noise. At $\sim 25$ days after the explosion (i.e., about a week after the $B$-band maximum), the P$_{\alpha}$ and P$_{\beta}$ signals are about 25\% level. Later on at $\sim 39$ and $58$ days after the explosion ($\sim 3$ and $6$ weeks after the $B$-band maximum), the signals reach 50-60\% level. This behavior is seen in all of our models (while it might be overestimated for the MS model). Note that while the epochs mentioned above are about one month since the $B$-band maximum, these are around (or just a bit later than) the NIR second-maximum date, i.e., much earlier than the previously proposed test for H$_{\alpha}$ emission in the late phase (about an year since the explosion).

H$_{\alpha}$ is indeed almost invisible to eyes at this flux level (Figure 10). In this sense, P$_{\beta}$ is more promising. Figure 18 shows that this feature could be visible even by inspections by eyes. Note that the flux-axis scale in this figure is reduced -- for example, Figure 6 shows P$_{\beta}$ more clearly. Figure 18 shows that the appearance of this feature is also dependent on the viewing angle. At $\sim 40$ days (around the NIR maximum), the feature is only visible for $\theta=0$. Later on at $\sim 60$ days the feature is visible for all $\theta$, with the larger flux and broader feature to the blue (i.e., larger velocity) for $\theta=0$. 

\begin{figure*}
\begin{center}
        \begin{minipage}[]{0.95\textwidth}
                \epsscale{1.0}
                \plotone{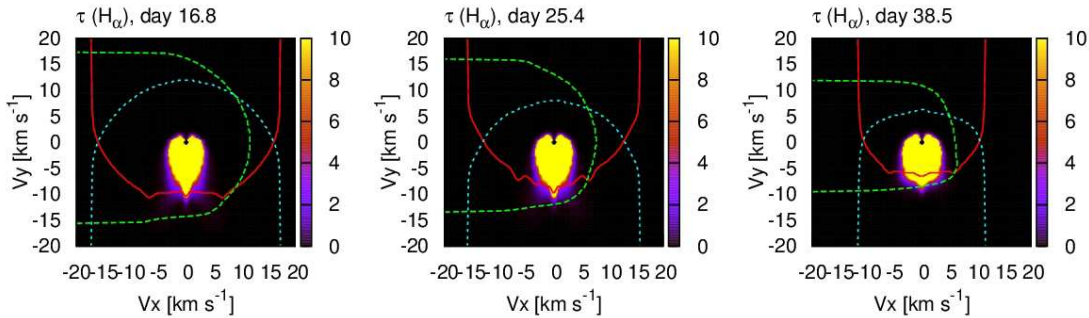}
        \end{minipage}
\end{center}
\caption
{Sobolev optical depth of H$_{\alpha}$ for Model RGa. Also shown here are the $R$-band photospheres for observers at different directions ($\tau = 2/3$; see the caption of Figure 7). 
\label{fig19}}
\end{figure*}

\begin{figure*}
\begin{center}
        \begin{minipage}[]{0.95\textwidth}
                \epsscale{1.0}
                \plotone{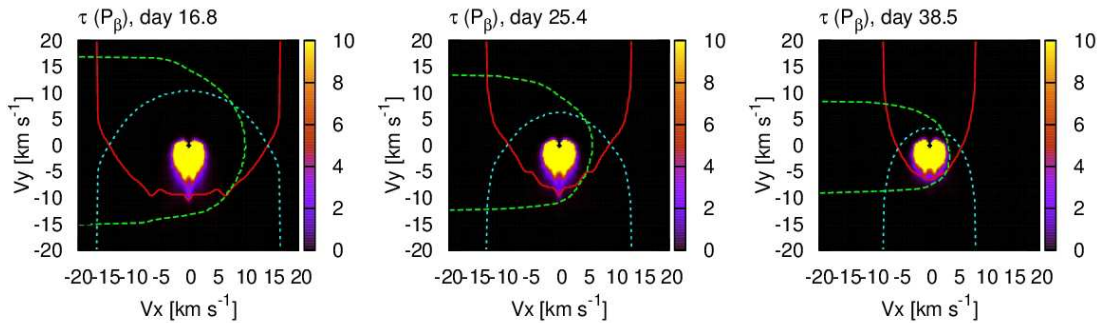}
        \end{minipage}
\end{center}
\caption
{Sobolev optical depth of P$_{\beta}$ for Model RGa. Also shown here are the $J$-band photospheres for observers at different directions ($\tau = 2/3$; see the caption of Figure 7).
\label{fig20}}
\end{figure*}

Figures 19 and 20 show the distribution of the Sobolev optical depths of H$_{\alpha}$ and P$_{\beta}$, respectively. Also shown are the underlying photosphere at $R$ and $J$-bands, respectively. Generally, the underlying optical depth is smaller in NIR than in optical, thus the H-rich region appears above the photosphere earlier in NIR. At day 38.5, hydrogen above the photosphere is mostly recombined (Fig. 8), and this explains why P$_{\beta}$ becomes stronger around this epoch. At this epoch, the photosphere as seen from an observer with the viewing angle of $\theta=\pi$ does not reach this neutral hydrogen region, thus at day 38.5, P$_{\beta}$ is visible only for an observer at $\theta=0$. The Sobolev optical depth is higher for H$_{\alpha}$ than P$_{\beta}$ and thus the self-absorption is more important in H$_{\alpha}$. This is one reason why P$_{\beta}$ is stronger than H$_{\alpha}$. 

As we see, in principle one could see hydrogen lines, especially $P_{\alpha}$ and P$_{\beta}$, {\em if one knows the hydrogen-free reference spectra a priori}. This is observationally a big issue in two respects. (1) One does not know the fraction of SN ejecta with hydrogen (through the interaction) -- this is what we aim to investigate. (2) When a given wavelength region is contaminated by other lines, there could be diversity (by some mechanisms) that is not directly related to the existence of hydrogen. For the point (2), even {\em assuming} that the observed SNe Ia are all represented by a series of models presented in this paper, the contaminating lines could show diversity arising from different viewing directions -- for example, H$_{\alpha}$ is contaminated by the much stronger Si line, and this line does show the diversity according to the viewing direction within our model (Figures 9 \& 10). This happens irrespective of the existence of hydrogen, even though in this model both the asymmetry and hydrogen contamination have the same origin. 

If there is a model that would perfectly describe (fit) the SN spectra one could rely on such model spectra. Unfortunately, this is not the case -- NLTE effects are suggested to become strong in post-maximum spectral formation especially in NIR, and so far there are no very `good' model spectra for it \citep[e.g., ][]{gall2012} -- while the models generally reproduce main features (including our synthetic spectra), the flux ratios of different lines can be strongly dependent on the NLTE treatment. Thus, to confirm the hydrogen lines observationally, one first has to define a `reference' (or template) spectrum from observed samples, and then one has to see the diversity of individual SN spectra as compared to the reference spectrum. The reference spectrum can be created either from the entire sample or a sub-set of the sample. For example, if one creates the reference spectrum from SNe with similar peak luminosity (or light curve width), that would effectively reduce a diversity related to the peak luminosity \citep[i.e., SN Ia spectral features generally correlate with the peak luminosity: ][]{nugent1995}. The remaining diversity may come from a combination of different effects \citep[see, e.g., ][]{maeda2010a,maeda2011} -- a strategy to detect hydrogen would be to see if there is a diversity associated with the wavelengths of the hydrogen lines. 

\begin{figure*}
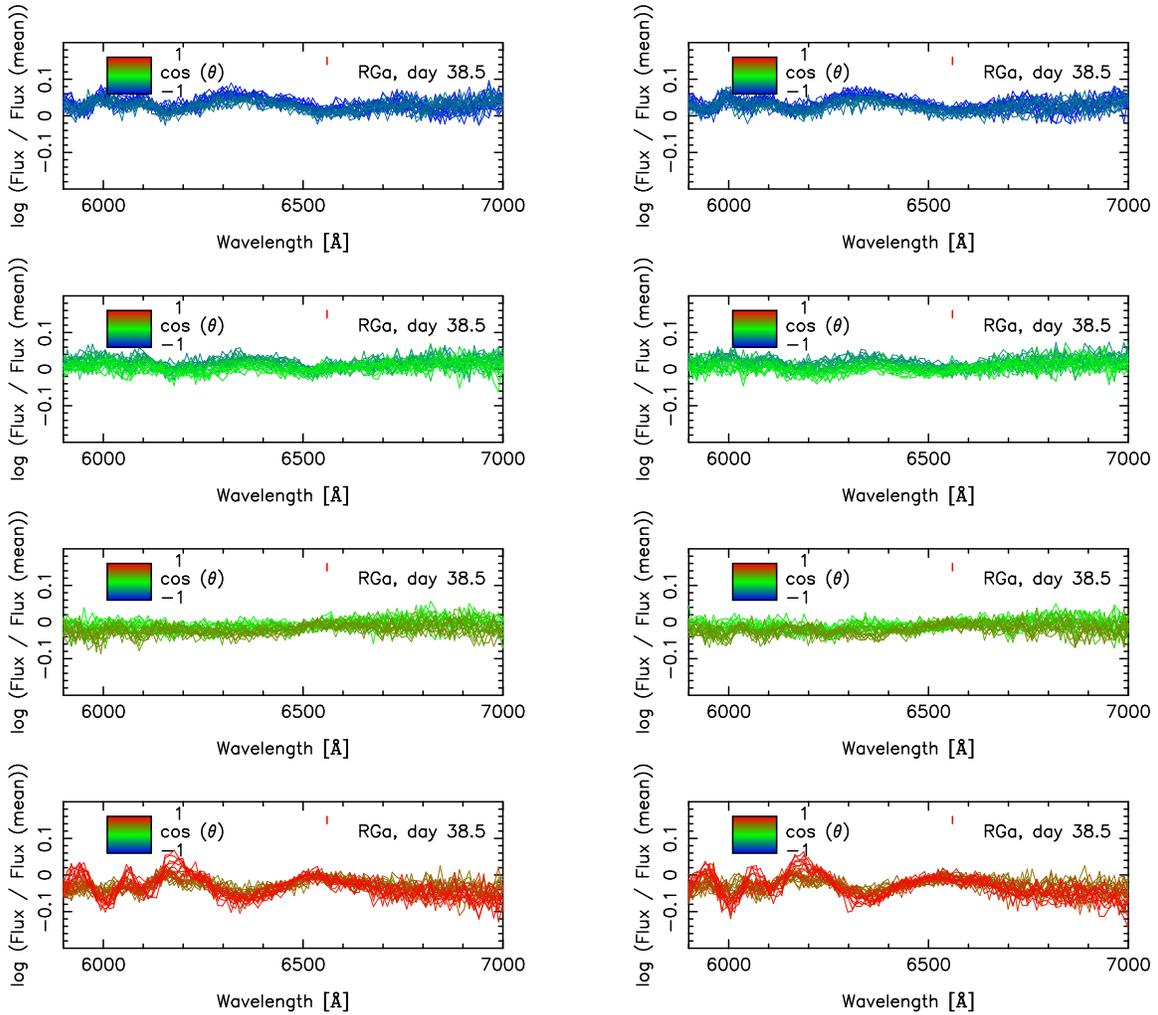

\begin{center}
        \begin{minipage}[]{0.45\textwidth}
                \epsscale{1.0}
                \plotone{f21a.eps}
        \end{minipage}
        \begin{minipage}[]{0.45\textwidth}
                \epsscale{1.0}
                \plotone{f21b.eps}
        \end{minipage}
        \begin{minipage}[]{0.45\textwidth}
                \epsscale{1.0}
                \plotone{f21c.eps}
        \end{minipage}
        \begin{minipage}[]{0.45\textwidth}
                \epsscale{1.0}
                \plotone{f21d.eps}
        \end{minipage}
        \begin{minipage}[]{0.45\textwidth}
                \epsscale{1.0}
                \plotone{f21e.eps}
        \end{minipage}
        \begin{minipage}[]{0.45\textwidth}
                \epsscale{1.0}
                \plotone{f21f.eps}
        \end{minipage}
        \begin{minipage}[]{0.45\textwidth}
                \epsscale{1.0}
                \plotone{f21g.eps}
        \end{minipage}
        \begin{minipage}[]{0.45\textwidth}
                \epsscale{1.0}
                \plotone{f21h.eps}
        \end{minipage}
\end{center}
\caption
{The residual of the synthetic spectra for Model RGa after being divided by the mean spectrum, shown for the optical range covering Si II 6355 and H$_{\alpha}$. The left panels are for the original model, while the right panels are for spectra artificially removing the hydrogen transitions. The panels are divided into four according to the viewing direction ($\theta = 0$ to $\theta = \pi$, from bottom to top). The synthetic spectra are binned within 3 time bins, but no additional binning is performed in the wavelength and viewing angle directions. The left and right panels are almost identical, showing that it is not observationally feasible to detect H$_{\alpha}$ at this epoch. Alternatively, the signature of overall ejecta asymmetry could be probed. 
\label{fig21}}
\end{figure*}

\begin{figure*}
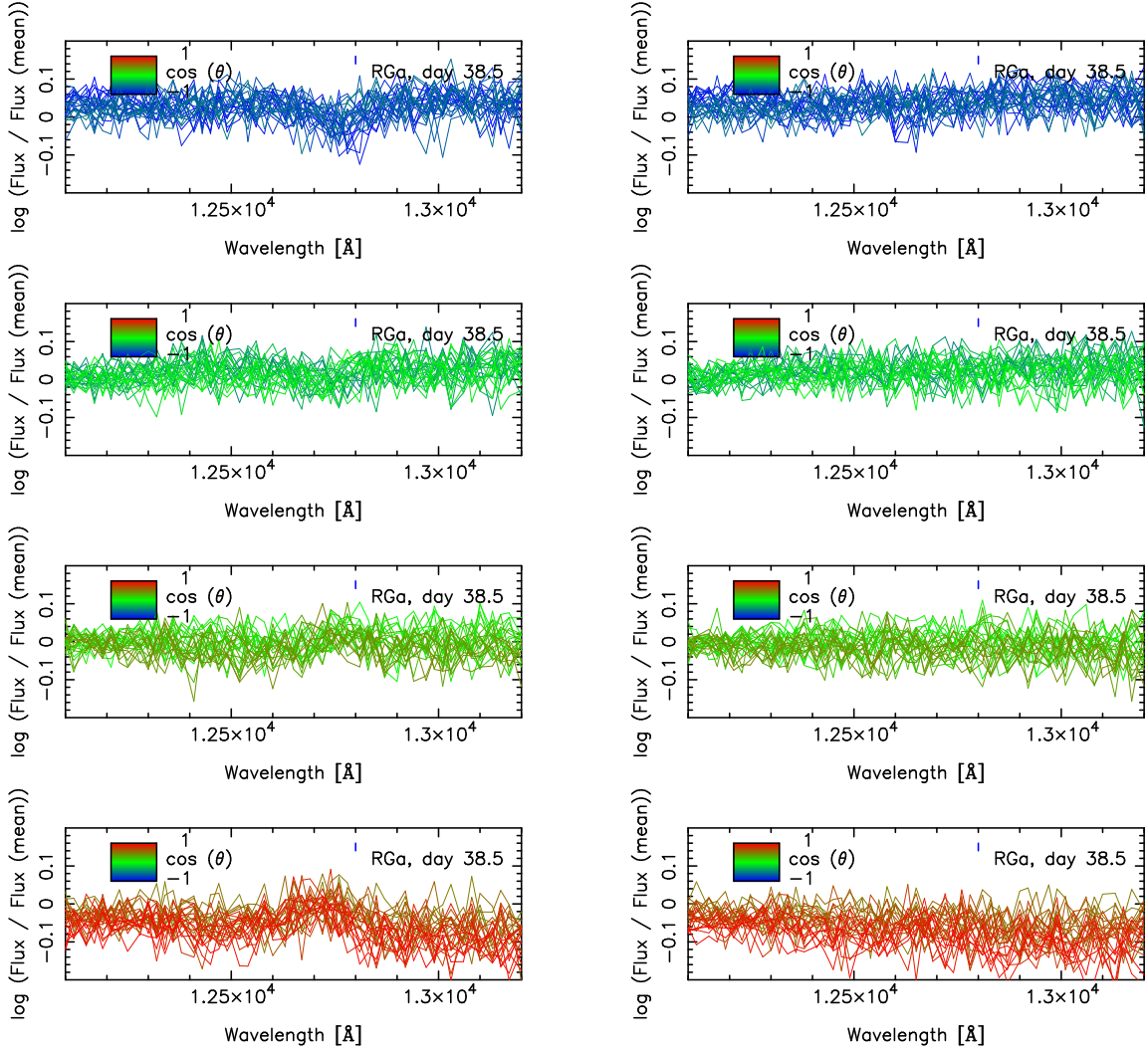

\begin{center}
        \begin{minipage}[]{0.45\textwidth}
                \epsscale{1.0}
                \plotone{f22a.eps}
        \end{minipage}
        \begin{minipage}[]{0.45\textwidth}
                \epsscale{1.0}
                \plotone{f22b.eps}
        \end{minipage}
        \begin{minipage}[]{0.45\textwidth}
                \epsscale{1.0}
                \plotone{f22c.eps}
        \end{minipage}
        \begin{minipage}[]{0.45\textwidth}
                \epsscale{1.0}
                \plotone{f22d.eps}
        \end{minipage}
        \begin{minipage}[]{0.45\textwidth}
                \epsscale{1.0}
                \plotone{f22e.eps}
        \end{minipage}
        \begin{minipage}[]{0.45\textwidth}
                \epsscale{1.0}
                \plotone{f22f.eps}
        \end{minipage}
        \begin{minipage}[]{0.45\textwidth}
                \epsscale{1.0}
                \plotone{f22g.eps}
        \end{minipage}
        \begin{minipage}[]{0.45\textwidth}
                \epsscale{1.0}
                \plotone{f22h.eps}
        \end{minipage}
\end{center}
\caption
{The residual of the synthetic spectra for Model RGa after being divided by the mean spectrum, shown for the NIR range covering P$_{\beta}$. The left panels are for the original model, while the right panels are for spectra artificially removing the hydrogen transitions. See the caption of Figure 21. The left panel shows the variation for different viewing directions due to P$_{\beta}$, while the model without hydrogen results in virtually no variation (right panel). As such, a possible diversity in the $J$-band spectra in many SN samples could be used to investigate the presence of hydrogen and a companion. 
\label{fig22}}
\end{figure*}

As an experiment, we here follow the same procedure as mentioned above to investigate if the hydrogen lines could be detectable based on Model RGa beyond other sources of diversities. We compare two cases, with and without hydrogen. In each case, we adopt an angle-averaged spectrum as a reference spectrum (at each epoch), and then compute a ratio of a spectrum viewed from a specific direction and the angle-averaged reference spectrum across the wavelength. This corresponds to the ratio of `individual' SN spectrum and the `reference' spectrum, but using the model spectra. Figure 21 shows results of this experiment for Model RGa with hydrogen (left panels) and without hydrogen (right panels), around the H$_{\alpha}$ at day 38.5. Figure 22 shows the same but for P$_{\beta}$. 

In general the optical range shows a larger diversity than NIR, due to the viewing direction difference. This highlights more complicated spectrum formation in optical (due to many overlapping lines) than in NIR at least at this epoch. This makes the identification of H$_{\alpha}$ quite difficult -- Namely, the `diversity' patterns with and without hydrogen are virtually identical, thus practically one cannot identify H$_{\alpha}$. On the other hand, P$_{\beta}$ shows a clear feature at its characteristics wavelength for the hydrogen-contaminated model spectra -- the model predicts that either absorption or emission could be seen at the wavelength of P$_{\beta}$ depending on the viewing direction. This feature is not seen in the case of the hydrogen-free model spectra. Thus, our model predicts that the diversity at the wavelength of P$_{\beta}$ should arise for model RGa, and for individual SNe it can either appear as emission or absorption depending on the viewing direction, when compared with a mean template spectrum. 

NIR spectra at this epoch are still rare, but the quick development of NIR detectors and increasing opportunities in NIR observations \citep{marion2006,marion2009} make it appealing, potentially powerful, diagnostics. For example, the magnitude of SN Ia 2003du was about $R \sim 20$ at 200 days since the explosion and $R \sim 22$ at one year after the explosion. It was $J \sim 16$ at $\sim 2$ weeks since the $B$-band maximum \citep{stanishev2007}, and must have been even brighter in $J$ at about one month after the $B$-maximum if we apply the Hsiao template light curve \citep{hsiao2007} (see also Appendix A). Thus, observational requirements for the `post-maximum' P$_{\beta}$ diagnostics may well be less tight than that for the `late-time' H$_{\alpha}$ diagnostics, in terms of the baseline sensitivity in different band passes.

\section{Conclusions and Discussion}
In this paper, we have investigated possible observational signatures of a non-degenerate companion star in the progenitor system of SNe Ia. Based on hydrodynamic simulations of the impact between expanding SN ejecta and the companion star, we have performed detailed radiation transfer simulations. We have focused on the maximum and post-maximum phases (covering the first two months) -- While the best data set of SNe Ia is available for these phases, the issue has been so far investigated mostly for the very early phase and late phase for which the observations are more challenging. 

\begin{figure*}
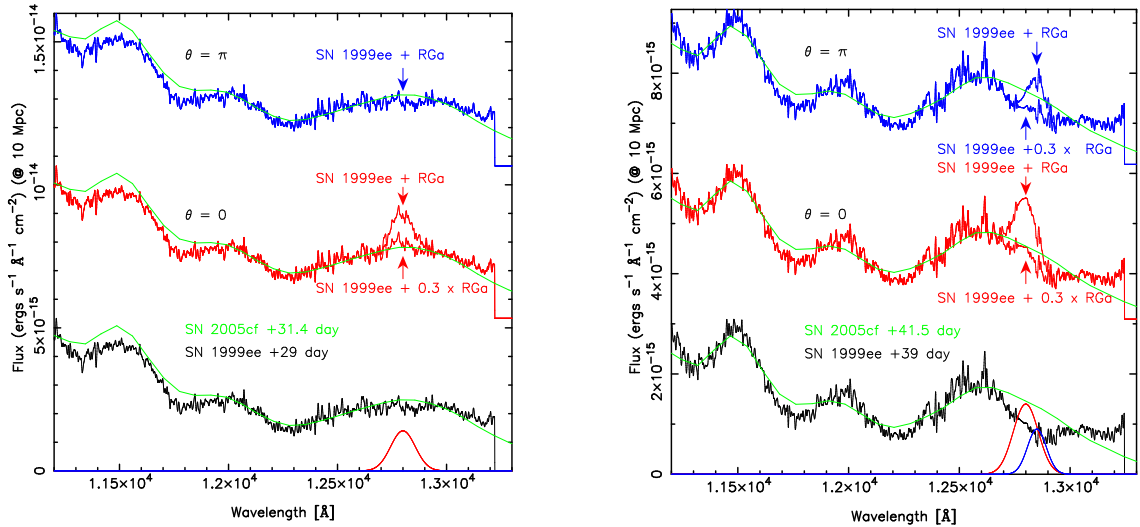

\begin{center}
        \begin{minipage}[]{0.45\textwidth}
                \epsscale{1.0}
                \plotone{f23a.eps}
        \end{minipage}
        \begin{minipage}[]{0.45\textwidth}
                \epsscale{1.0}
                \plotone{f23b.eps}
        \end{minipage}
\end{center}
\caption
{Observed $J$-band spectra (bottom) of SNe 1999ee (black) and 2005cf (green) at $\sim 30$ days since the $B$-band maximum (left) and $\sim 40$ days (right). The flux of SN 2005cf is brought to the hypothesized distance of 10 Mpc, assuming the original distance of 28 Mpc. The flux of SN 1999ee is scaled to roughly fit the flux of 2005cf at a similar epoch in the $J$-band. The P$_{\beta}$ line in the synthetic spectra is approximated by a Gaussian profile, extracted from Model RGa (bottom, red for $\theta = 0$ and blue for $\theta = \pi$). This is added to the spectrum of SN 1999ee for an observer at $\theta = 0$ (middle) and at $\theta = \pi$ (top). In doing this, two cases are shown (see the labels in the figure) -- one with the original prediction and the one where the synthetic P$_{\beta}$ flux is multiplied by 0.3 (roughly corresponding to the H-rich envelope mass scaled down to $0.1 M_{\odot}$). This kind of analysis could be used to constrain the amount of stripped-off hydrogen, thus the existence of a companion star (see main text for details). 
\label{fig23}}
\end{figure*}

Compared to a previous study by \citet{kasen2004}, our approach is different in the following aspects: (1) we start with the hydrodynamic simulations rather than assuming a simplified kinetic model, (2) we follow the temporal evolution, (3) we analyze not only the optical properties but also NIR, with an extended analysis of various observationally testable features. Because of the differences, our model predictions indeed differ from those by \citet{kasen2004} even qualitatively. 

We have found that the overall properties, especially photometric ones, are not much different between the systems with and without a companion star, even with the RG companion. Interestingly, we find in our simulations that the light curves seen from the companion side are not bluer and brighter as suggested in the previous study \citep{kasen2004}, but we predict the opposite. The difference is however generally at a 0.1 magnitude level. Therefore, the existence of a non-degenerate companion star is not ruled out for individual SNe, by the currently available maximum and post-maximum data of SNe Ia. The model predicts a diversity arising from different viewing angle (at the level of 0.1 magnitude), showing some correlations between different colors and magnitudes. In the optical wavelength band, interestingly the expected relations are similar to those introduced by an external extinction, whose nature is yet to be clarified. This indicates that this effect, if the progenitor with a non-degenerate companion explains a good fraction of observed SNe Ia, can introduce systematic errors at the level of 0.1 magnitude in using SNe Ia as standard candles. We have found that the NIR properties do not follow the external-extinction properties, highlighting the importance of NIR observations in developing the SN Ia luminosity/distance calibration better than the 0.1 magnitude level. 

The difference between models with and without a companion is bigger in spectroscopic features than in photometric features. We predict that the Si II 6355 velocity (and other lines) depends on the viewing direction. At the maximum brightness, the Si II 6355 velocity is smaller for an observer viewing from the companion side ($\theta=0$), as is consistent with the result by \citet{kasen2004}. The temporal evolution of the feature shows an even more interesting behavior. Before the maximum, the Si II feature does not show a strong viewing angle dependence. Later on, since just before the maximum date, the Si II velocity starts decreasing quickly for an observer viewing from the companion side. The predicted relations between the velocity and the optical ($B-V$) color, as well as the velocity and the `velocity gradient' are found to be different than those inferred from the observations. Thus, the angle variation on the companion-induced asymmetry cannot be a source of these relations. Indeed, we do not try to explain the relations but alternatively suggest to use these relations to constrain the existence of a non-degenerate companion star. There should be other mechanisms (e.g., the mass of $^{56}$Ni and other factors) that introduce the observed diversity/relation, thus the variation due to the viewing angle based on the present model should be regarded to be the `minimum' variation. Therefore the predicted variations should not be larger than the observational variations. If this is violated, it means that such a model does not account for a bulk of the observed SNe Ia. Comparing the predicted variations with the observed scatters in the velocity--color and velocity -- velocity gradient relations, we have found that the model is marginally consistent with the current observations. In the future, observations with better calibrations (especially in photometry) is expected to place a strong constraint on the existence of a non-degenerate companion star from this aspect. 

We have also investigated if there is a chance to probe the non-degenerate companion through the hydrogen features in maximum and post-maximum spectra. We confirmed the expectation (while not quantitatively shown before) that H$_{\alpha}$ is difficult (practically impossible) to detect in these phases. Alternatively, we suggest that P$_{\beta}$ can potentially be used as a diagnostics around the NIR maximum phase (or slightly later). We have shown that detecting this feature is observationally feasible, and can be even easier than the search of the H$_{\alpha}$ emission in the later phases.

As a demonstration of the observational feasibility, in Figure 23 we show a comparison between NIR spectra of two SNe Ia, 1999ee \citep{hamuy2002} and 2005cf \citep{gall2012}, which have the published NIR data at similar epochs. The data were obtained through the Weizmann interactive supernova data repository \citep{yaron2012}. The comparison shows that these two SNe are extremely similar in NIR, and this similarity in NIR provides an ideal situation to investigate particular features (in this case P$_{\beta}$) since defining the `continuum' (or template) is relatively straightforward. There is indeed a hint of the developing difference around P$_{\beta}$ between the two SNe in the later phase ($\sim 40$ days since the $B$-band maximum) while one has to carefully check the data reduction process to confirm it is not an artifact. This is beyond the scope of this paper, and we will examine a sample of NIR data in a separate paper (K. Maeda, in prep.). 

Here as a demonstration we use the spectra of SN 1999ee as templates, and investigate the constraint on the amount of hydrogen contaminated in the ejecta of SN 2005cf. Figure 23 also shows the $P_{\beta}$ at the corresponding epochs, shown in the bottom of both panels, extracted from Model RGa for $\theta = 0$ (red) and $\pi$ (blue). This flux is then added to the original (observed) spectrum of SN 1999ee. This way, we can check if the contamination of the H-rich materials in SN 2005cf is consistent with the model, assuming that there is no H-rich material contaminated in SN 1999ee. Further, by varying the model flux, we can place a constraint on the amount of H allowed for SN 2005cf.

At $\sim 30$ days since the $B$-band maximum, Model RGa predicts that P$_{\beta}$ is visible if viewed from the companion side ($\theta \sim 0$) while not so from the opposite side ($\theta \sim \pi$). Adding the P$_{\beta}$ flux predicted by Model RGa for an observer at $\theta = 0$ to the spectrum of SN 1999ee, this exceeds the observed flux of SN 2005cf at that wavelength. Thus, a situation that SN 2005cf had a RG companion in the close binary system {\em and} it was viewed from the companion side is ruled out. If one reduces the predicted P$_{\beta}$ flux to 30\% of the original (corresponding to $\sim 0.1 M_{\odot}$ of the mixed hydrogen), this would not conflict with the observation. Any companion star is not ruled out for an observer at $\theta=\pi$. Later on at $\sim 40$ days since the $B$-band maximum, a similar but tighter constraint can be obtained: $M$(H) $\lsim 0.1 M_{\odot}$ for an observer at $\theta = 0$ and $\lsim 0.2 M_{\odot}$ for $\theta = \pi$. Note that the SN ejecta contaminated with $0.1 M_{\odot}$ of hydrogen can be a typical feature of the SN ejecta-companion impact \citep[see, e.g., ][]{liu2012}, thus the diagnostics we propose here can be quite powerful to identify/rule out a non-degenerate companion star. 

Note also that by comparing two SNe, we indeed constrain {\em difference} in the hydrogen content in these two SNe. Thus, it is necessary to construct a `hydrogen-free' template spectrum from a large sample. In doing this, there are several possibilities in the template construction. Dividing the SN sample into subgroups with different peak luminosities (or decline rate) is an obvious choice. One would also be tempted to divide the sample according to the host types or environment properties, then compare the templates for each group as well as compare individual SNe with the templates. Such a strategy may pick up possible different populations in different environments \citep[e.g., ][]{wang2013} providing a test of how different populations may be related to the SD and DD scenarios. 

There are a few limitations in the present study. We have adopted the expansion opacity formalism and the two-level atom approximation, rather than simulating the full details of the fluorescence following the excitations. This is a good approximation for ions with complicated level structures like Fe-peaks, since the high-rate of the radiation-matter interactions should establish a quasi-equilibrium which is represented by thermal redistribution \citep[see, e.g., \S 3.6 of ][]{kasen2006}. The use of this prescription for a simple atom like hydrogen may need further justification and calibration. 

Another issue is on the NLTE effects. For SNe Ia, the NLTE effect on ionization is especially strong at $\gsim 50$ days since the explosion while it is not so in the earlier phase \citep{kromer2009}. Around the peak date, a strong effect can be seen in UV, while the effect is not significant in the optical range if an appropriate value for the thermalization parameter is adopted \citep{baron1996}. For SNe IIp, which could be relevant to our investigating the hydrogen lines, deviations from LTE have significant effects on line profiles, while the continuum flux is not much affected \citep{baron1996,dessart2008,kasen2009} (ses, e.g., Fig. 3 of \citet{baron1996}). Indeed, a bigger effect is expected for {\em time-dependent} NLTE effects, which for SNe IIp models could introduce a change in the flux level of H$_{\alpha}$ by a factor of a few \citep{dessart2008}. Still, this effect would not remove all the H$_{\alpha}$ flux predicted in LTE calculations \citep{dessart2008}, and thus we would not expect that our results will be changed qualitatively due to the NLTE effects. 

Related to the NLTE effects, in our formalism the effect of the fluorescence is taken into account by a single thermalization parameter ($\epsilon$), and it has been shown that the biggest difference from the present prescription ($\epsilon=0.3$) is expected for the pure scattering atmosphere ($\epsilon=0$) \citep{baron1996,kasen2006}. To check if this particular choice of $\epsilon$ (calibrated for metal lines) would affect the strength of hydrogen features, we have repeated the same calculations for Model RGa but setting $\epsilon = 0$.  The result is shown in Appendix C. We conclude that this does not introduce much difference. We caution that the two-level approximation and the thermalization parameter cannot be exactly calibrated to a full NLTE description, therefore introducing the thermalization parameter is merely an approximation. Ultimately it should be tested by the full NLTE calculations for the same models. Still, as mentioned above, the results from the previous studies on SN Ia and IIp models, both relevant to our study, are promising, suggesting that this approximation would not introduce significant errors in the observables of interest in this paper.

We note that neglecting non-thermal excitations of hydrogen by $\gamma$-rays might indeed lead to underestimates of the hydrogen line fluxes. Also, in our model we omit a metal content in the companion envelope, by assuming the purely hydrogen and helium in it. This would not change the overall feature, since the main part of the emission is created by the `SN ejecta' and the companion envelope merely dilutes the emission through Thomson scattering. An inclusion of the metal would increase the heating of the H-rich region, therefore would keep the ionization of hydrogen high for longer time than in our present model. However, as time goes by the $\gamma-ray$ heating, which is not sensitive to the metal content, becomes progressively important. As a result, our prediction of the appearance of P$_{\beta}$ in relatively late phases would not be dramatically affected by the metal content in the companion envelope. 

Besides the hydrogen issue, the ejecta asymmetry is a characteristic feature of a non-degenerate companion system, and we predict that this configuration leads to a characteristic `diversity pattern' across the wavelength (Figs. 21 \& 22). In Appendix D we show the expected diversity patterns at different epochs for model RGa. In the same way as we propose for searching P$_{\beta}$, the comparison of a spectrum of an individual SN and a template spectrum can in principle be used to search for such a diversity pattern. This is another way we propose to search for a signature of non-degenerate companion stars in SN Ia systems.

\acknowledgements 
This research is partly supported by World Premier International Research Center
Initiative (WPI Initiative), MEXT, Japan. K. M. acknowledges financial support by Grant-in-Aid for Scientific Research for young Scientists (23740141, 26800100). Numerical computations presented in this paper were carried out using 512 cores on Cray XC30 at Center for Computational Astrophysics, National Astronomical Observatory of Japan. We thank Ken'ichi Nomoto for providing the W7 model and constructive discussion, and Elena Sorokina and Sergei Blinnikov for providing their synthetic light curves for the W7 model. We also thank the anonymous referee for her/his extensive review and many constructive comments. 
We have used the Weizmann interactive supernova data repository (www.weizmann.ac.il/astrophysics/wiserep) to obtain the archival spectra data of SNe 1999ee and 2005cf.

\appendix
\section{A. Method of Radiation Transfer}

Our radiation transfer simulation code adopts the Monte Carlo method, where paths of individual photon packets are computed as a random walk process. This method is broadly adopted in radiation transfer simulations in SN ejecta \citep[][and references therein]{lucy2005}. It is suited to treat Doppler shifts of photons due to the velocity gradient in the expanding/moving medium and resulting enhancement of the bound-bound opacities or `expansion opacities' \citep{karp1977,eastman1993} -- due to the successive Doppler shifts in a comoving frame a photon experiences as it flies through the moving medium, it can suffer from the discrete transitions (bound-bound) at frequencies different from the original frequency of a photon at its creation. 

We largely adopt prescriptions given by \citet{lucy2005}, \citet{kasen2006}, and \citet{kromer2009}. Our simulation is purely radiation transfer, thus the kinetic and composition structure [$\vec v (\vec r, t), \rho (\vec r, t), X_{i} (\vec r, t)$] should be provided as a background (i.e., no feedback process from the radiation to hydrodynamics is taken into account). The radiation transfer simulation provides iteratively the thermal and ionization conditions and accordingly the distribution of opacities [$T (\vec r, t), n_{i}^{j} (\vec r, t), \alpha_{\lambda} (\vec r, t)$] (where $n_{i}^{j}$ is the number density of an ion $j$ at $i$-th level) so as to be consistent with the radiation field [$f_{\lambda} (\vec r, t, \vec l)$] (here $\vec l$ is the photon direction vector) under the assumptions of {\em LTE} and radiative equilibrium. 

The code is applicable for the 1D spherical coordinate, the 2D spherical-polar coordinate, and the 3D spherical-polar and Cartesian coordinates, which can be simply specified in an input parameter file. The 1D and 2D versions assume spherical symmetry and axisymmetry so that the number of photon packets can be reduced to reach to the convergence. We have performed various test calculations to confirm that the imposed asymmetry does not introduce errors in the transfer simulations (an example is described later). 

The multi-dimensional/frequency/epoch radiation transfer code, {\em HEIMDALL (Handling Emission In Multi-Dimension for spectrAL and Light curve calculations)}, takes the following steps in simulating the radiation transfer. The main part of the code is written in a general way so that the applicability is not restricted to the radiation transfer in the SN ejecta, but below when necessary specific functions for the SN radiation transfer are described. The code is written in a hybrid-parallelization mode using {\em openMP} and {\em MPI} and its parallelization efficiency has been tested up to 512 cores distributed over 64 cpus. 

{\bf 1: Determining the distribution of initial photon packets: } \\
Over the course of the main MC routine, the properties of a photon packet are described by a set of variables [$\vec r (t), \vec l (t), \lambda (t), \varepsilon (t)$], i.e., the position, direction vector, wavelength, and the total energy within the packet. Here, we describe the photon packet (or a MC quanta) as a group of identical photons (or particles), i.e., $\varepsilon = n_{\rm ph} h \nu$ where $n_{\rm ph}$ is the number of the photons in a packet and $\nu$ is the frequency. To compute the change in these variables as a function of time ($t$) by the main MC routine, we have to determine the initial conditions, i.e., [$\vec r (t_0), \vec l (t_0), \lambda (t_0), \varepsilon (t_0)$] where $t_0$ is the time of the creation of the thermal photon under consideration.  

Photon packets created by processes other than interactions of already existing thermal photons and matter are specified at the beginning of simulations (note that photons created by such interactions between already existing thermal photons and matter are treated over the course of the main MC transfer). For simulations performed in this paper, these are photons created as a result of radioactive decay energy input through the decay chain $^{56}$Ni $\to$ $^{56}$Co $\to$ $^{56}$Fe. These $\gamma$-ray and positron packets are created at the beginning of a simulation based on the distribution of $^{56}$Ni and its decay property -- the photon packets are assigned with the spectral energy (i.e., branching ratios in the decays) and time of the creation (i.e., decay time) determined with random number generation. If the time at the creation of a $\gamma$-ray packet, as determined by the MC random number generation for every packet, is earlier than the starting time of the whole simulation, the $\gamma$-ray packet is assumed to be absorbed by the simulation start time at the position of the creation in the comoving frame. This deposited energy is converted to optical photons at the starting time of the optical photon transfer, taking into account the adiabatic loss of the thermal energy between the deposition time and the simulation starting time, assuming that the optical photons created here have diffused negligibly to matter in this time interval \citep{lucy2005,kromer2009}. This treatment is justified by the short mean free paths of photons in the early phase, while the approximation becomes less robust if the starting time for the transfer simulation is taken to be later. Our simulations are started at 10 days after the explosion (approximately a week before the $B$-band maximum), and we have checked the applicability of this approximation in Appendix B, where we find that our results are not sensitive to this relatively late starting time of the simulations. Transfer of $\gamma$-rays are solved taking into account Compton scattering, pair creation, and photoelectric absorption based on the scheme identical to optical photons but without temperature iteration since the cross sections of these interactions are insensitive to the thermal condition. Positrons are assumed to deposit their energy in situ. The details of the computational method here are given by \citet{maeda2006a}. During the MC transfer simulation, the energy deposition by $\gamma$-rays and positrons, $\Gamma_{\gamma} (\vec r, t)$ are tracked as in the same manner with the heating by UV/optical photons (see below). 

Since the energy deposition by $\gamma$-rays and positrons is the only source of the thermal energy (i.e., ultimately the energy of the thermal radiation) in the present situation, the transfer simulation described above provides the initial condition for the thermal photon packets to be followed by the main MC routines. With the energy deposition rate $\Gamma_{\gamma} (\vec r, t)$ we thus determine the total energy content of the thermal photon packets emitted at a given spatial bin and a time bin. In the calculations shown in this paper, the energy content of each thermal photon packet ($\varepsilon$) is set to be equal for all the packets at its creation. We first integrate the energy deposition rate in space and time, then the number of thermal photons created at a given spatial bin and a given time bin is determined by the relative contribution of $\Gamma_{\gamma} (\vec r, t)$ (as integrated within a spatial bin and time bin) to the total deposited energy. The position and time at its creation within the spatial and time bins are determined with the random number generation (in the comoving frame). The direction vector of the packet is also computed by the random number generation assuming the isotropic emission in the comoving frame. Now we thus have a set of variables to specify the properties of the photon packets in the comoving frame, except for its wavelength, and these are transformed to the SN-rest (or observer) frame. The wavelength, $\lambda$ ($t_0$), cannot be specified at this step, as it requires the temperature to be known. Thus, the creation of the photon packet is coupled with the main MC routine and iteratively solved following the steps described below, to be self-consistent with the ejecta temperature. 

{\bf 2: Computing thermal and ionization structures and opacity distributions: } \\
At given time and for given temperature $T (\vec r, t)$, the ionization and level populations are computed under the {\em LTE} assumption, i.e., through the Saha equation and the Boltzmann distribution. Then, the opacities are computed as a function of wavelength, including bound-bound, bound-free, free-free, and electron scattering. The electron scattering opacity is computed with the electron number density, $n_{\rm e} (\vec r, t)$, which is  given by the ionization condition: 
\begin{equation}
\alpha_{\rm e} (\vec r, t) = \sigma_{\rm T} \ n_{\rm e} (\vec r, t), 
\end{equation}
where $\sigma_{\rm T}$ is the Thomson cross section. The free-free absorption opacity is computed as follows: 
\begin{equation}
\alpha_{\rm ff} (\vec r, t, \nu) = \frac{4 e^6}{3 m_{e} h c} \left(\frac{2 \pi}{3 k m_e}\right)^{1/2} T (\vec r, t)^{-1/2} \sum_{j} Z_{j}^2 
n_{\rm e} n^j \nu^{-3} \left[1 - \exp\left(\frac{h \nu}{k T}\right)\right] g_{\rm ff} \ , 
\end{equation}
where $Z_{j}$ is the number of free electrons associated with the ion $j$, and $n^j$ is the number density of the ion $j$. 
The Gaunt factor ($g_{\rm ff}$) is set to be unity. For the bound-free transitions, cross sections [$\alpha_{\rm ff} (\vec r, t, \lambda)$] are taken from \citet{verner1996} and \citet{verner1996}. For given $\lambda$, the bound-free cross sections are summed over for different ions (with the ionization states determined by the Saha equation). 

The bound-bound transitions are treated in the Sobolev approximation, where the `line' optical depth is given as follows: 
\begin{equation}
\tau_{lu} (\vec r, t)= \frac{\pi e^2}{m_e c} \ f_{lu} \ \lambda_{lu} \ t \ n_{l} (\vec r, t) \left[1 - \frac{g_l n_u (\vec r, t)}{g_u n_l (\vec r, t)}\right] \ , 
\end{equation}
where subscripts $l$ and $u$ denote the lower and upper levels of a transition under consideration (here we omit the superscript $j$ to specify the ion). $f_{lu}$ and $\lambda_{lu}$ are the oscillator strength and the wavelength of the transition. $g_{l}$ and $g_{u}$ are statistical weights of the lower and upper level, respectively. For the line list, we adopt a standard set of $5 \times 10^5$ bound-bound transitions from \citet{kurucz1995}. 

With the Sobolev optical depth, the escape probability of the photon out of the resonance region is 
\begin{equation}
\beta_{lu} (\vec r, t) = \frac{1 - e^{-\tau_{lu} (\vec r, t)}}{\tau_{lu} (\vec r, t)} \ .
\end{equation}
We treat the bound-bound transitions within the expansion opacity formalism, i.e., combining the transitions into a discrete frequency grid \citep{karp1977,eastman1993}. Here, the total cross section at wavelength $\lambda$ is given as 
\begin{equation}
\alpha_{\rm bb} (\vec r, t, \lambda) = \frac{1}{c t} \sum_{l, u} \frac{\lambda_{lu}}{\Delta\lambda} (1 - e^{-\tau_{lu}}) \ ,
\end{equation}
where the sum runs over the bound-bound transitions whose energy difference is within the wavelength bin under consideration ($\Delta\lambda$). 
The purely absorptive component is defined within the two-level atom approximation, i.e., 
\begin{equation}
S_{\lambda} = (1 - \epsilon_{lu}) J_{\lambda} + \epsilon_{lu} B_\lambda (T) \ , 
\end{equation}   
where the source function is divided into the scattering component (i.e., treated as a resonance line) and into the absorptive component (i.e., thermalized after multiple scatterings and fluorescence). Generally, this treatment of the bound-bound transitions is shown to provide a good approximation for the thermal conditions appropriate to SNe Ia, since the large opacities and many transitions lead to the thermal redistribution. This has been calibrated with more detailed transfer where the fluorescence is directly treated \citep{baron1996,kasen2006}. While $\epsilon_{lu}$ is dependent on different transitions, the result of the radiation transfer is indeed insensitive to the exact value of $\epsilon_{lu}$ as long as $\epsilon_{lu} \sim 1$ \citep[see, e.g., Fig. 10 of][]{kasen2006}. For this reason, we adopt the same value for all the transitions, $\epsilon \equiv \epsilon_{lu} = 0.3$ as our standard case, following \citet{kasen2006}. We do caution that the argument is dependent on the focus and topics under investigation. For example, the direct treatment of the fluorescence is essential in the late-time spectral formation \citep{kromer2009}. 
Now, the explicit form for the purely absorptive component in the bound-bound transitions can be written as follows: 
\begin{equation}
\alpha_{\rm bb, abs} (\vec r, t, \lambda) = \frac{1}{c t} \sum_{l, u} \frac{\lambda_{lu}}{\Delta\lambda} \frac{\epsilon_{lu}}{\beta_{lu}+\epsilon_{lu} (1 - \beta_{lu})} (1 - e^{-\tau_{lu}}) \ .
\end{equation}

The total opacity is given as the sum of the different components described above: 
\begin{equation}
\alpha (\vec r, t, \lambda) = \alpha_{\rm e} (\vec r, t) + \alpha_{\rm ff} (\vec r, t, \lambda) + \alpha_{\rm bf} (\vec r, t, \lambda) + \alpha_{\rm bb} (\vec r, t, \lambda) \ .
\end{equation}
The purely absorptive component is defined as follows: 
\begin{equation}
\alpha_{\rm abs} (\vec r, t, \lambda) = \alpha_{\rm ff} (\vec r, t, \lambda) + \alpha_{\rm bf} (\vec r, t, \lambda) + \alpha_{\rm bb, abs} (\vec r, t, \lambda) \ .
\end{equation}

{\bf 3: Propagation of photon packets through the MC simulation: }\\ 
With the background condition including opacity distribution (in space and frequency) now specified, the propagation of photon packets is computed from time $t_{n}$ to $t_{n+1}$. When the photon packet already exists at $t_{n}$ from the previous time step, its path until $t_{n+1}$ or until it escapes out of the ejecta before $t_{n+1}$ is followed by the MC simulation for each photon packet, and this procedure is repeated for all the photon packets. If the photon packet is created at $t_0$ between $t_{n}$ and $t_{n+1}$ (due to the $\gamma$-ray and positron deposition) its path is followed from $t_0$. 

At each step, photon path lengths for several numerical and physical events are computed, then the event with the minimal length is adopted as the real event. These events include the following cases. (1) A photon reaches to a boundary between the current spatial grid and one of the neighboring grids. (2) The physical time the packet experiences reaches to the next time step ($t_{n+1}$), (3) The photon comoving spectral frequency is redshifted to come into the next frequency bin, (4) A photon suffers from either scattering or absorption. This procedure is repeated for all the photon packets. 

The item (3) is specific for the radiation transfer in moving medium like the SN ejecta. Since we assume the homologous expansion, the Doppler shift is simply computed as $\Delta\lambda = \lambda v/c$, and this inversely gives the distance the packet travels before suffering from the Doppler shift of the amount $\Delta\lambda$. The item (4) is evaluated through the standard MC formula as follows: 
\begin{equation}
\alpha' (\vec r, t, \lambda) \rho (\vec r, t) \delta s' = -\ln z \ ,
\end{equation}
where $z$ is the random number (between 0 and 1) and $\delta s$ is the path length. Here the prime indicates the rest-frame quantities. 

When the packet experiences the interaction, its fate after the interaction is again determined through the random number generation, proportional to cross sections to each event. Specifically, we judge if this is a scattering or absorption, according to the ratio of $\alpha_{\rm abs} (\vec r, t, \lambda)$ and $\alpha_{\rm scat} \equiv \alpha (\vec r, t, \lambda) - \alpha_{\rm abs} (\vec r, t, \lambda)$. 

A scattering is treated as an isotropic and coherent scattering in the comoving frame, a good approximation for resonance transitions (and Thomson scattering). A new photon direction in the comoving frame is chosen randomly following a standard MC procedure and it is transferred to the rest frame. It is elastic in the comoving frame, and the transfer from the comoving frame to the rest frame automatically takes into account the adiabatic loss in a microscopic sense. 
\begin{figure*}
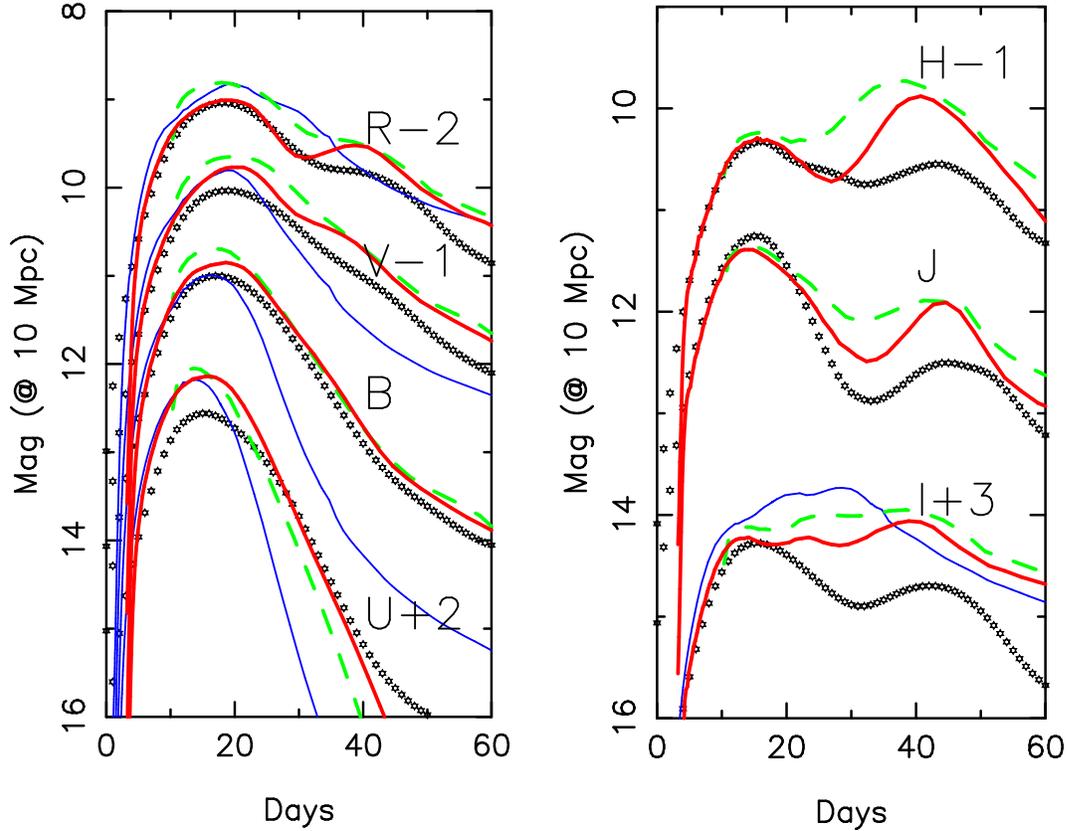


\begin{center}
\vspace{5cm}
\hspace{2.0cm}
        \begin{minipage}[]{0.45\textwidth}
                \epsscale{1.6}
                \plotone{fa1a.eps}
        \end{minipage}
\hspace{-1cm}
        \begin{minipage}[]{0.45\textwidth}
                \epsscale{1.6}
                \plotone{fa1b.eps}
        \end{minipage}
\end{center}
\caption
{Synthetic multi-band light curves as compared with SN Ia template light curves. The SN Ia template light curves (stars) are constructed from the Hsiao spectral template \citep{hsiao2007} convolved with standard filter functions. Our synthetic light curves based on W7 model are shown by red curves (thick solid). For comparison, the synthetic light curves computed by STELLA are shown by blue curves (thin solid), from U to I band. Our 2D `reference' W7 model is shown by green curves (thick dashed). For the `reference' model, we apply an offset of 0.34 magnitudes for all the bands for fair comparison, since the model has a larger amount of $^{56}$Ni (see the main text). The amount of the offset here reflects the difference in the mass of $^{56}$Ni ($0.81 M_{\odot}$ in the `reference' model and $0.59 M_{\odot}$ in the original W7 model). 
\label{figa1}}
\end{figure*}

\begin{figure*}
\begin{center}
\vspace{7cm}
\hspace{12cm}
        \begin{minipage}[]{0.95\textwidth}
                \epsscale{1.0}
                \plotone{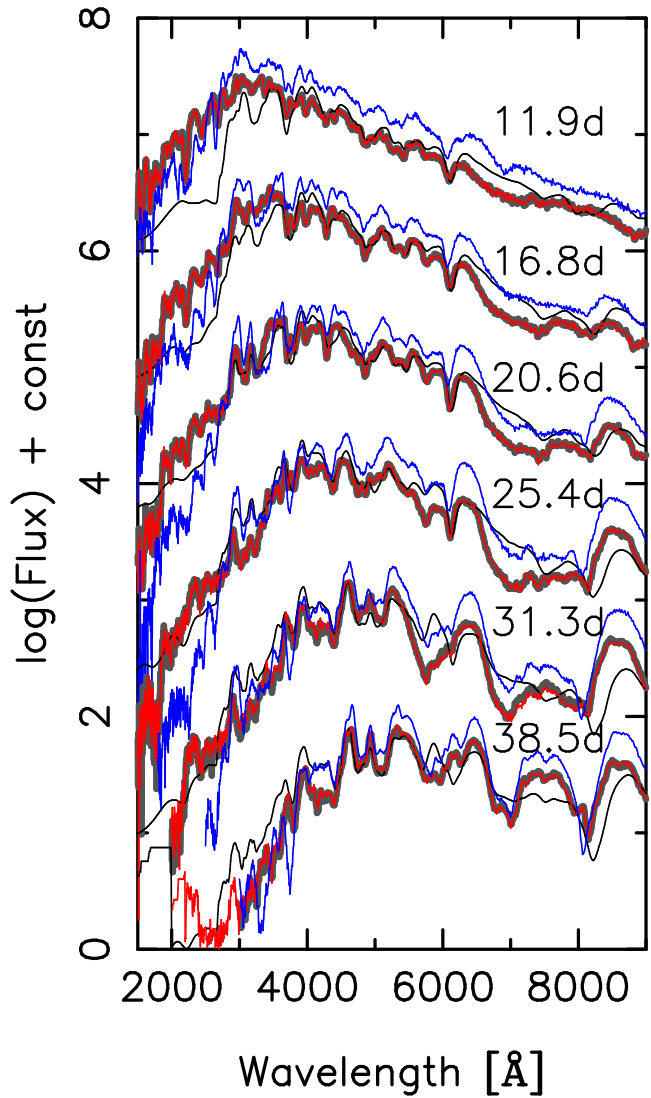}
        \end{minipage}
\end{center}
\caption
{Synthetic spectra as compared with the Hsiao SN Ia template. The zero-point in the epoch is the explosion time, and the $B$-maximum date is assumed to be 17 days since the explosion to label the Hsiao template spectra. The W7 model spectra (as computed under the assumption of spherically symmetry) are shown by red, while the 2D `reference' model spectra are shown by blue. The Hsiao template spectra are shown by black. Synthetic spectra of W7, but computed with the radiation transfer scheme in 2D space under the assumption of axisymmetry is shown by (thick) gray lines, to show that our scheme of the radiation transfer in 2D does not introduce any artifact. The values of the offset applied to each spectra are the same for all the models (i.e., no additional offset is applied to provide the best match between the models and the templates). For the spectra at late epochs, a blue portion of the spectra is truncated in the presentation, where the MC noise is large due to the small amount of UV photons in the late phase. 
\label{figa2}}
\end{figure*}

An absorption and re-emission is treated within the thermalization approximation, and thus a new wavelength at its re-emission is determined through the local thermal emissivity: 
\begin{equation}
j_\lambda (\vec r, t) = B_\lambda(T) \alpha_{\rm abs} (\vec r, t, \lambda) \ .
\end{equation}
Here the emission is treated as isotropic in the comoving frame. 

Note that the propagation is treated in the rest frame, using the cross sections originally computed in the comoving frame but transferred into the rest frame. The physical events (scattering and absorption/re-emission) are treated in the comoving frame, i.e., first the rest frame quantities are transferred into the comoving frame, computing the outcome in the comoving frame, and then the result is transferred back into the rest frame. The formalisms for the transformation are given by \citet{castor1972}. This way, the Doppler shift in the moving medium is appropriately handled, and for example it results in the P-Cygni profile for bound-bound transitions. 

{\bf 4: New Temperature determination and iteration for the temperature convergence: }\\
Over the course of the photon propagation, the heating rate between $t_{n}$ to $t_{n+1}$ at each spatial grid is tracked, using the MC estimator: 
\begin{equation}
\Gamma_{\rm opt} (\vec r, t) = \frac{1}{\Delta t V} \sum_{k} \alpha_{\rm abs} (\vec r, t, \lambda) \varepsilon_{k} \delta s_{k} \ ,
\end{equation}
where the quantities are given in the comoving frame, and this estimator runs over all the packet (as specified by $k$) which passes through a given grid between $t_{n}$ and $t_{n+1}$ ($\Delta t \equiv t_{n+1} - t_{n}$). Here, $V$ is the volume of the spatial grid. With the heating rate by $\gamma$-rays and positrons obtained in the same manner ($\Gamma_{\gamma}$), the heating-cooling balance under the radiative equilibrium provides a constraint on the temperature, 
\begin{equation}
\Lambda (T) = \Gamma_{\rm opt} (\vec r, t)+ \Gamma_{\gamma} (\vec r, t) \ ,
\end{equation}
where the cooling rate at each spatial grid is given as 
\begin{equation}
\Lambda (T) = 4 \pi \int \alpha_{\lambda} (T) B_{\lambda} (T) d\lambda \ ,
\end{equation}
where the absorptive opacity $\alpha_{\lambda} (T)$ is approximated by the one estimated with the `previous' temperature [i.e., $\alpha_{\rm abs} (\vec r, t, \lambda)$]. These equations give new temperature estimate.

The steps 2 - 4 are repeated for a given time step (between $t_{n}$ and $t_{n+1}$) until the temperature converges at all the meshes simultaneously. Once this happens, the converged temperature is used for the initial guess for the temperature in the next time step, and the photon packets' properties are used as the initial conditions for the propagation calculations in the next time step (at $t_{n+1}$). Then, the procedures 2 - 4 are repeated in the next time step until the temperature convergence. This way, we proceed with time, following the radiation field and thermal condition in a self-consistent manner.

{\bf 5: Extraction of synthetic spectra and light curves: }\\
In the MC packet propagation routine, the paths of every photon packet are followed. When the photon packets escape out of the SN ejecta (or the numerical domain), the information is recorded. This provides the escaping radiation flux as a function of the viewing direction, time, and wavelength [$f_{\lambda} (\vec l, t)$] (here $\vec l$ is the photon direction vector). From this we extract angle-dependent spectra as a function of time. The light curves in multi band passes are then extracted by convolving the filter functions to the synthetic spectral sequence. In this paper, we use the Johnson and Kron-Cousins systems for UBVRI and 2MASS system for NIR.

Figure A1 shows an example of the synthetic light curves for the W7 model. We find a reasonable agreement between our result and a result obtained by an independent simulation code {\em STELLA} \citep{blinnikov1998,blinnikov2006}. While different codes generally agree to reproduce overall behaviors, details are different depending on specific treatments \citep[see, e.g., ][]{kromer2009}. Our results are well within these variations, and similar to the result of \citet{kasen2006}. We find a reasonable agreement between the W7 model prediction and the Hsiao template light curves \citep{hsiao2007}. The discrepancy is larger in NIR than in optical, but this is a general issue in the radiation transfer simulations for SNe Ia \citep[see, e.g., ][]{kromer2009,gall2012}. We note that `our reference' model (the modified W7) also shows a behavior similar to the original W7, justifying to use this model as our reference model.

Figure A2 shows a synthetic spectral sequence for the W7 model (red; computed in 1D under the assumption of spherical symmetry), the `reference' model (blue), and the Hsiao template spectra (black) \citep{hsiao2007}. We see a reasonable agreement between the W7 model and the Hsiao templates. There are deviations especially in the later phases -- while the model does predict spectral features at correct wavelengths, the strengths of the features can be different from the observed templates. This is suggested to be caused by {\em NLTE} effects \citep[e.g.,][]{kromer2009}, and is a generic issue for spectrum synthesis in SNe \citep[see, e.g.,][for 3D delayed-detonation models]{sim2013}. Our `reference' model shows a larger flux than the W7 and the Hsiao templates due to the large amount of $^{56}$Ni, but otherwise the predicted features are very similar to the W7 model. Therefore, using this model as our reference is justified. We also show the same W7 models but mapped onto the 2D grids and computed in 2D, under the assumption of axisymmetry (but no symmetry with respect to the equatorial direction). We see a perfect match between the 1D and 2D calculations, proving that our 2D radiation transfer scheme does not introduce any errors in the transfer simulation. 

\section{B. Robustness of the Predicted Maximum-Phase Behaviors}

In this paper, we deal with the diversity in the magnitudes and colors around the $B$-band maximum at 0.1 magnitude level. In this section, we show that our simulations are accurate to this level to claim the diversity arising from the different viewing directions. Especially, we address the following two points: (1) If the relatively late starting time in our simulations (10 days) affects the claimed behaviors, and (2) if the predicted diversities and correlations are not affected by possible numerical instabilities. 

\begin{figure*}
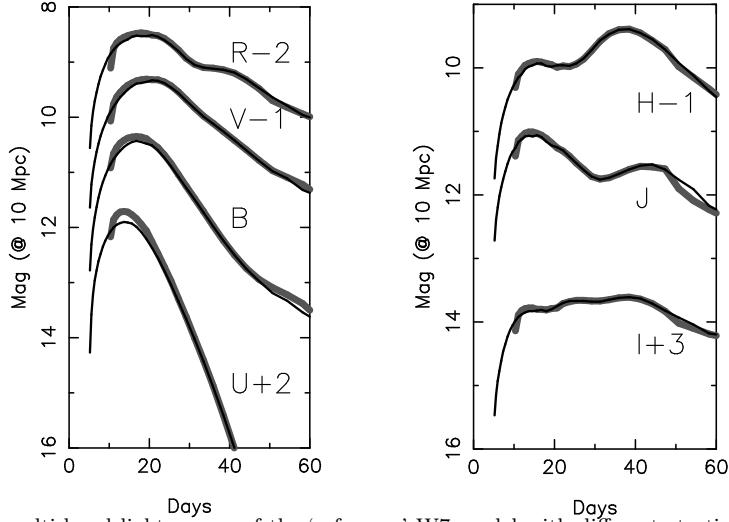

\begin{center}
\vspace{2.0cm}
\hspace{1.0cm}
        \begin{minipage}[]{0.45\textwidth}
                \epsscale{1.0}
                \plotone{fb1a.eps}
        \end{minipage}
        \begin{minipage}[]{0.45\textwidth}
                \epsscale{1.0}
                \plotone{fb1b.eps}
        \end{minipage}
\end{center}
\caption
{Synthetic multi-band light curves of the `reference' W7 model with different starting time in the simulations. The original simulation (starting at day 10) is shown in gray, while the simulation starting at day 5 is shown in black. 
\label{figb1}}
\end{figure*}

\begin{figure*}
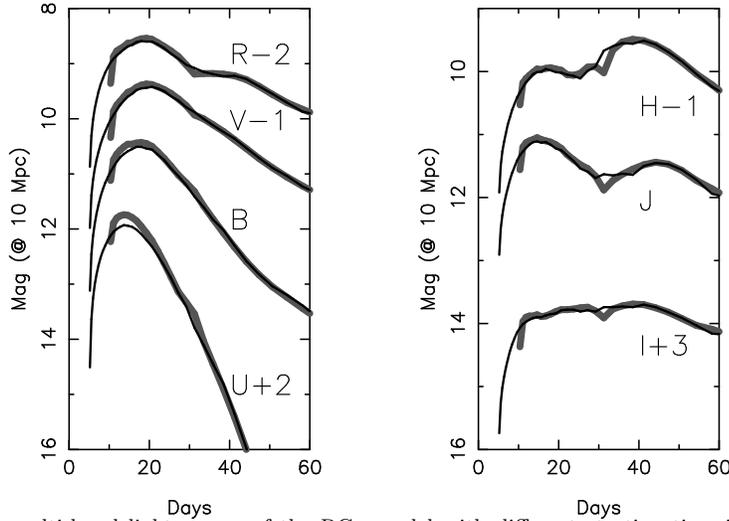

\begin{center}
\vspace{2.0cm}
\hspace{1.0cm}
        \begin{minipage}[]{0.45\textwidth}
                \epsscale{1.0}
                \plotone{fb2a.eps}
        \end{minipage}
        \begin{minipage}[]{0.45\textwidth}
                \epsscale{1.0}
                \plotone{fb2b.eps}
        \end{minipage}
\end{center}
\caption
{Synthetic multi-band light curves of the RGa model with different starting time in the simulations. Shown here are the angle-averaged mean light curves. The original simulation (starting at day 10) is shown in gray, while the simulation starting at day 5 is shown in black. 
\label{figb2}}
\end{figure*}

Figure B1 shows the multi-band light curves of the `reference' W7 model from the simulation starting at day 5, as compared to our standard run starting at day 10. The same but for the RGa model is shown in Figure B2. It is seen that the two calculations with different starting time converge quickly toward the $B$-band maximum date in both models. A substantial difference is seen in the $U$-band light curve around its peak date (before the $B$-band peak), while in the other bands the difference is small. This test also shows that (late-phase) kinks seen in the original calculations especially in the $J$-band (i.e., $\sim 50$ days for the reference model and $\sim 30$ days in the RGa model) are numerical artifacts. The fact that these kinks appear much later than the $B$-band peak where the treatment of the starting time should be unimportant indicates that this late-phase stability can be sensitive to small variation in the thermal conditions, but fortunately it seems that this unstable behavior does not appear around the maximum phase (see also below). 

\begin{figure*}
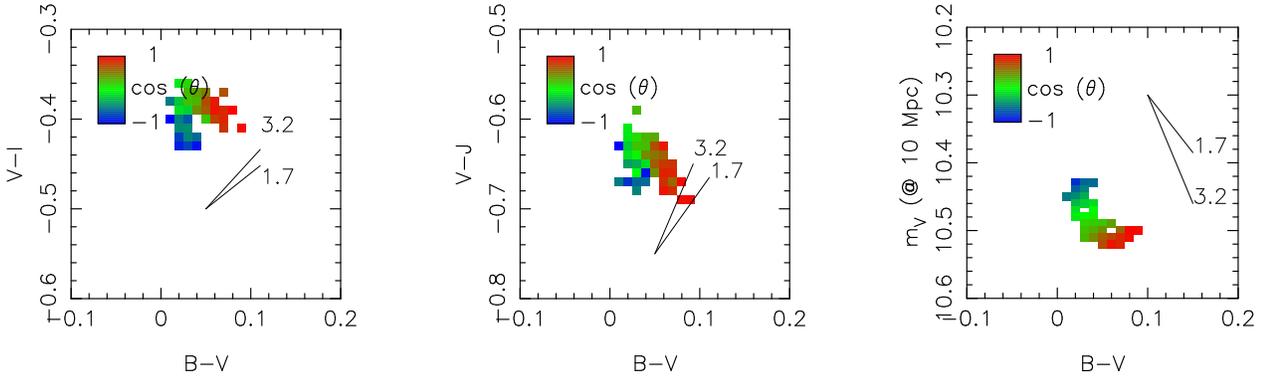

\begin{center}
        \begin{minipage}[]{0.32\textwidth}
                \epsscale{1.0}
                \plotone{fb3a.eps}
        \end{minipage}
        \begin{minipage}[]{0.32\textwidth}
                \epsscale{1.0}
                \plotone{fb3b.eps}
        \end{minipage}
        \begin{minipage}[]{0.32\textwidth}
                \epsscale{1.0}
                \plotone{fb3c.eps}
        \end{minipage}
\end{center}
\caption
{The predicted relations in photometric properties for Model RGa as same as Figure 12, but for the simulation starting at day 5. 
\label{figb3}}
\end{figure*}

Figure B3 shows the variations in the colors and the $V$-band magnitude, and their relations to the $B - V$ color obtained for Model RGa, by the simulation starting at day 5. This should be compared to Figure 12 where the simulation is started at day 10. While a small offset in the absolute scale is seen for the $B-V$ and $V-I$ colors and the $V$-band magnitude at the level of 0.05 magnitude, the trend as a function of different viewing directions and the amount of the resulting diversity are consistent with the original simulation. 

\vspace{5cm}
\begin{figure*}
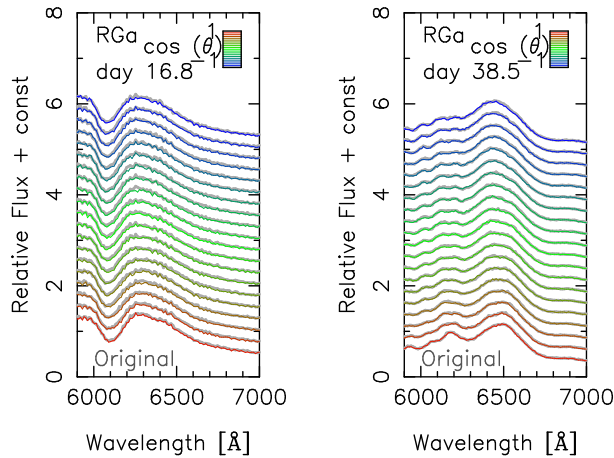

\begin{center}
\hspace{1cm}
        \begin{minipage}[]{0.23\textwidth}
                \epsscale{1.7}
                \plotone{fb4a.eps}
        \end{minipage}
        \begin{minipage}[]{0.23\textwidth}
                \epsscale{1.7}
                \plotone{fb4b.eps}
        \end{minipage}
\end{center}
\vspace{2cm}
\caption
{The simulated spectra around Si II 6355 (from the RGa model) as same as Figure 10 but for the simulation starting at day 5 (color curves). The color coordinates indicate the prediction for different viewing directions (red for $\theta = 0$ and blue for $\theta = \pi$). In this figure, the spectra for observers at different directions are added with an additional offset ($\theta = 0$ to $\pi$, from bottom to top). The gray curves here are the spectra obtained by the original simualtion starting at day 10. 
\label{figb4}}
\end{figure*}

The spectral features are even less sensitive to the starting time of the simulation. Figure B4 compares the spectra around Si II 6355 in both simulations. It is seen that the results are almost identical, and therefore the relations involving the Si velocity should also be unaffected by the treatment of the starting time.

We note that while the absolute magnitude can be affected by the treatment of the starting time (which turns out to be 0.05 mag level), the behavior in the colors and magnitudes arising from different viewing direction should be much less sensitive here. The spectral features are expected even less sensitive. These are confirmed by the test simulation, justifying our claims regarding the correlation and diversities arising from different viewing directions. 

Another issue is if the trend and diversity arising from the different viewing directions around the $B$-band maximum are affected by numerical instability similar to the one seen in the $J$-band light curves in the later phase. First, the investigation of the sensitivity on the starting time suggests that such an instability takes place less likely in the earlier phase than the later phase; the different thermal conditions due to the different starting time may mimic numerical instability in the thermal condition, but we see that the thermal condition quickly converges before the $B$-band maximum. This is expected, since the radiation--matter coupling is quite strong in the early phase which should suppress the numerical instability quickly (note that the iteration in the thermal condition is performed in every time step under the assumption of radiative equilibrium). Next, for such an instability to affect the {\em viewing-angle dependence}, the instability itself should create a strong angle-dependent effect, which should be seen as an {\em angle-dependent} sudden rise and fall in the multi-band light curves. Even for the possible instability found in the later phases (in the $J$-band), this effect does not show strong angle-dependence. Therefore, even if a similar numerical instability would take place in the earlier phase, it is unlikely that such a putative effect should affect the angle-dependent effects we claim in the present paper. 

To support these arguments, we have performed a test calculation based on the `reference' W7 model. Here, on day 13.6 (i.e., $\sim 3$ days before the $B$-band maximum), we artificially reduce the temperature in the ejecta only within $\theta = 165 - 180$ degree to the 40\% of the `real' converged temperature. As such, if the instability does not fade away, this should create artificially--introduced viewing--angle dependence for the expected observables. The setup here is chosen, following a trial and error, so that the inserted artificial effect on the viewing angle variation exceeds the variation seen in the RGa model. 

Figure B5 shows the variation of the $V$-band and $H$-band magnitudes in the RGa model due to different viewing directions, as compared to the angle-averaged mean magnitude in each band. The evolution for given $\theta$ is quite continuous, and we do not see any sudden change in the viewing angle dependence which would announce any possible numerical instabilities. This should be compared with Figure B6 which shows the same quantities for the reference W7 model where the numerical instability is artificially inserted in the cone (in the direction of $\theta = 165 - 180$ degree). We see here a sudden `broadening' of the angle dependence especially in the $H$-band. Such a behavior is not seen in the RGa model, suggesting that any instability which creates numerically-introduced viewing dependence larger than the `real' physical behavior is not present in the simulation for the RGa model. 

\begin{figure*}
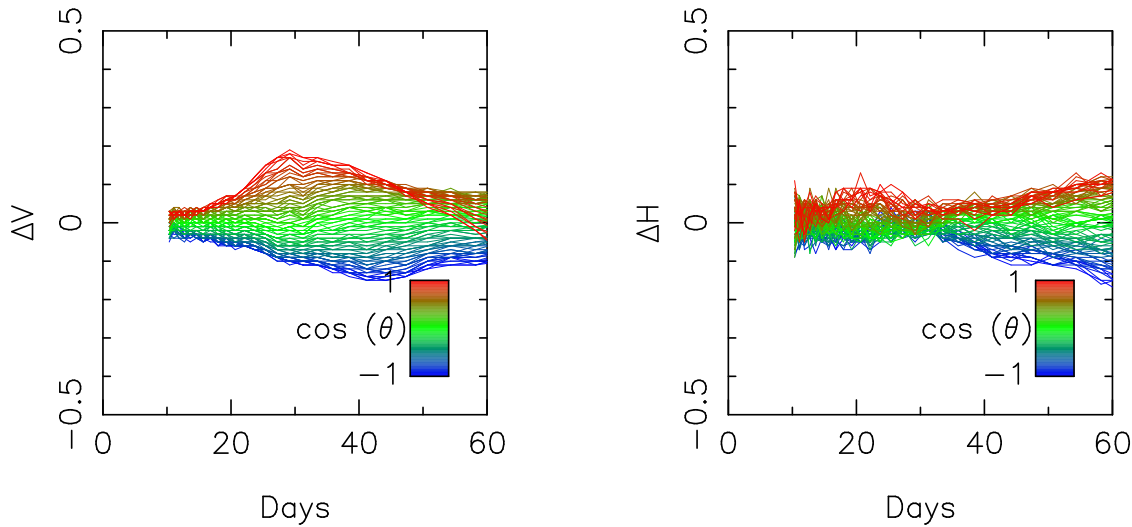

\begin{center}
        \begin{minipage}[]{0.45\textwidth}
                \epsscale{1.0}
                \plotone{fb5a.eps}
        \end{minipage}
        \begin{minipage}[]{0.45\textwidth}
                \epsscale{1.0}
                \plotone{fb5b.eps}
        \end{minipage}
\end{center}
\caption
{Evolution of the difference between the magnitude for observers at various directions and the angle-averaged magnitude for the RGa model. Shown here are the $V$-band and $H$-band magnitudes. 
\label{figb5}}
\end{figure*}

\begin{figure*}
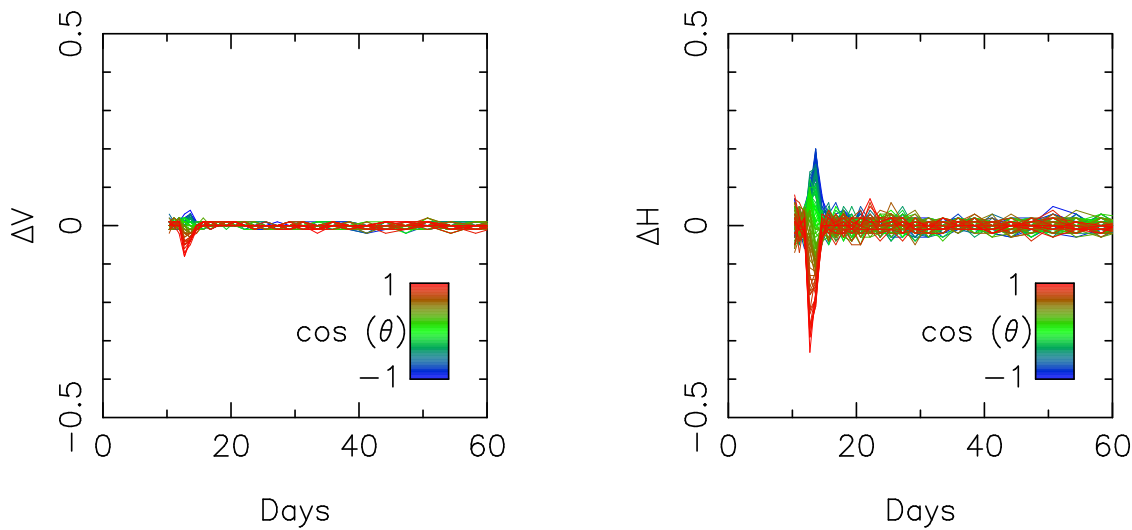

\begin{center}
        \begin{minipage}[]{0.45\textwidth}
                \epsscale{1.0}
                \plotone{fb6a.eps}
        \end{minipage}
        \begin{minipage}[]{0.45\textwidth}
                \epsscale{1.0}
                \plotone{fb6b.eps}
        \end{minipage}
\end{center}
\caption
{Evolution of the difference between the magnitude for observers at various directions and the angle-averaged magnitude for the `reference' W7 model but with the instability in the thermal condition artificially introduced at $\theta = 165 - 180$ degree on day 13.6 (i.e., $\sim 3$ days before the $B$-band maximum). Shown here are the $V$-band and $H$-band magnitudes. 
\label{figb6}}
\end{figure*}

Furthermore, the artificially--introduced variation for different viewing directions quickly fades away, and the remaining variation (due to the MC noise) becomes much smaller than the variations seen in the RGa model. This further supports that the viewing angle dependence around the $B$-band maximum is not affected by any instabilities, and the MC noise in our simulations is sufficiently small. In sum, we conclude that our claim on the viewing-angle dependence on the $B$-band maximum observables (e.g., Figures 12 and 14) are not due to numerical artifacts. 

\section{C. Treatment of Hydrogen lines}

\begin{figure*}
\begin{center}
        \begin{minipage}[]{0.95\textwidth}
                \epsscale{0.7}
                \plotone{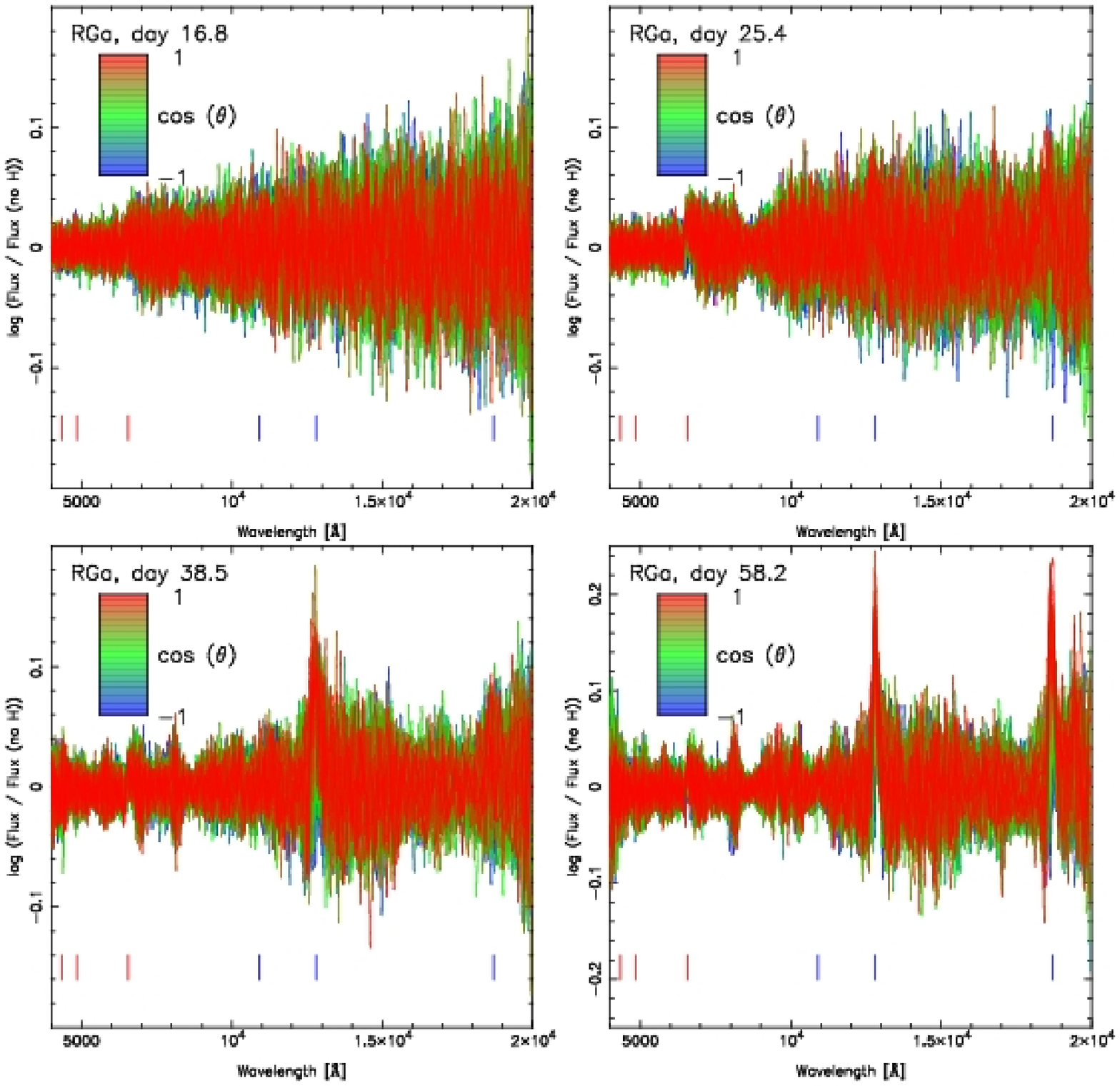}
        \end{minipage}
\end{center}
\caption
{The ratio of the spectral flux with and without hydrogen lines in Model RGa. This is the same with Figure 15, except that the hydrogen bound-bound transitions are treated as fully resonance lines (i.e., $\epsilon=0$). 
\label{figc1}}
\end{figure*}

As we have adopted the expansion opacity formalism and the two-level atom approximation, a question is if the predicted fluxes of the hydrogen features can be significantly affected by these assumptions (\S 6). Previous studies suggest that it would not introduce a large difference \citep{baron1996,dessart2008,kasen2009}. Here, we evaluate a possible effect of this for the situation we investigate. 

We are interested to see if the hydrogen line fluxes are much affected by the treatment of the bound-bound transitions for given thermal condition. Thus, we have computed the model spectra for Model RGa under an extreme assumption on the hydrogen line formation, namely purely scattering (fully resonance) line transitions within our formalism ($\epsilon=0$). To do this, we have taken the thermal condition computed from the original calculations ($\epsilon = 0.3$) so that we can purely pick up the effect of $\epsilon$ on the line formation. Then, the model calculations with and without hydrogen (for latter we artificially set the cross sections of hydrogen to zero) are performed, and the ratio is investigated as we did in \S 5. Figure C1 shows the result. We thereby confirmed that this does not affect the detectability of the hydrogen lines. As we expect that the scattering dominated atmosphere is an extreme condition, we believe that the hydrogen line formation we find in this paper would not be much affected by the treatment of hydrogen line transitions.

\section{D. Diversity Patterns from the companion-induced asymmetry}

Irrespective of the hydrogen content in the expanding SN ejecta, the companion-induced asymmetry can create characteristic diversity patterns as a function of the wavelength. This can in principle be tested by observations, by studying a diversity seen in spectra of individual SNe as compared to a reference/template spectrum. The reference spectrum should be created as a mean of the spectra of SNe in a specific sub-group (or all SNe Ia), and the comparison can be performed between an individual and the reference spectra. It will allow to see possible diversity of individual SNe from the average behavior of the group. The diversity pattern predicted from the companion-induced asymmetry is different for different viewing directions, and it evolves with time. A specific example of the expected diversity pattern is shown in Figures 21 and 22 for an epoch of 38.5 days since the explosion. In Figures D1--D3, we provide the model predictions for different epochs.

\begin{figure*}
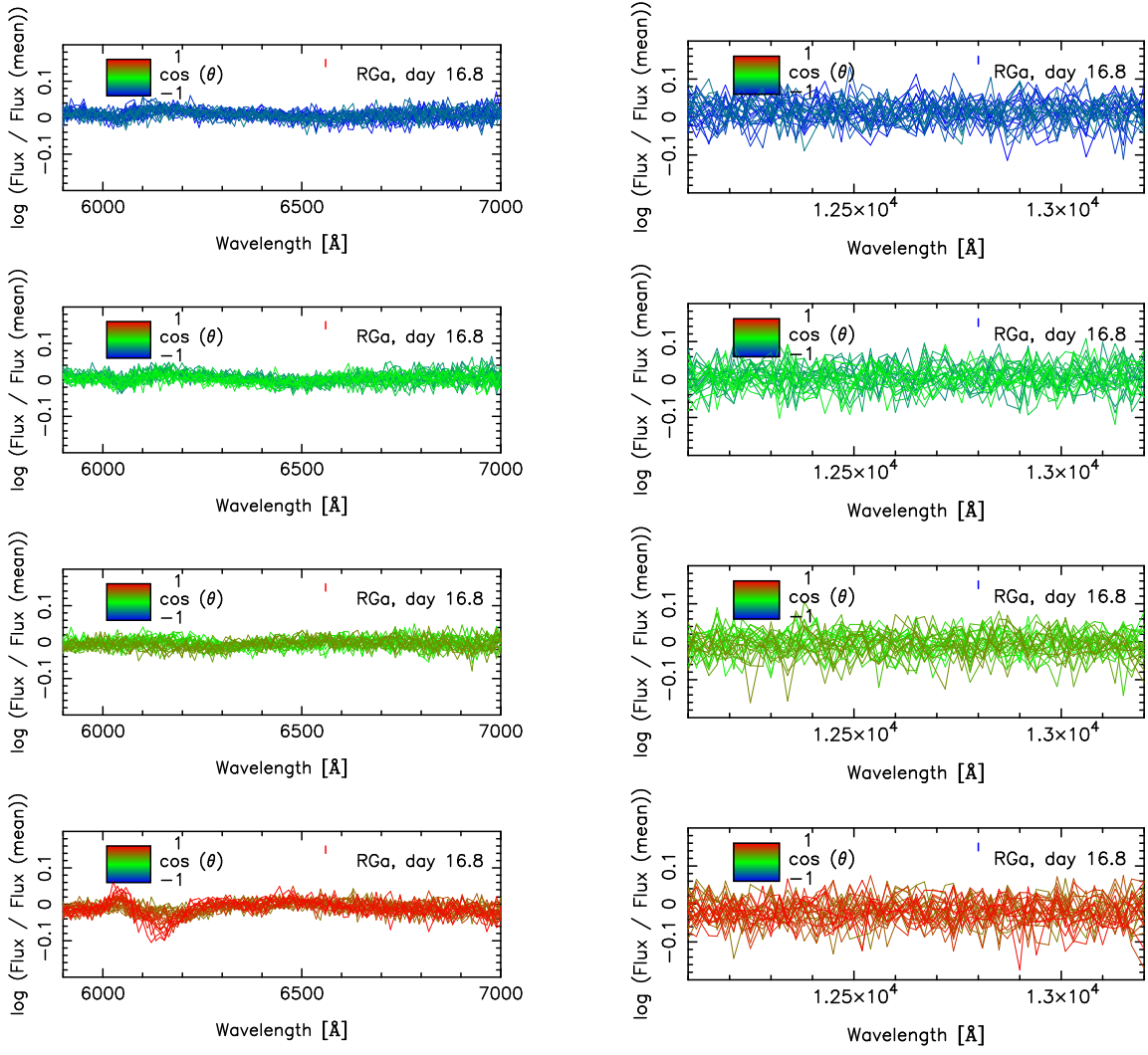

\begin{center}
        \begin{minipage}[]{0.45\textwidth}
                \epsscale{1.0}
                \plotone{fd1a.eps}
        \end{minipage}
        \begin{minipage}[]{0.45\textwidth}
                \epsscale{1.0}
                \plotone{fd1b.eps}
        \end{minipage}
        \begin{minipage}[]{0.45\textwidth}
                \epsscale{1.0}
                \plotone{fd1c.eps}
        \end{minipage}
        \begin{minipage}[]{0.45\textwidth}
                \epsscale{1.0}
                \plotone{fd1d.eps}
        \end{minipage}
        \begin{minipage}[]{0.45\textwidth}
                \epsscale{1.0}
                \plotone{fd1e.eps}
        \end{minipage}
        \begin{minipage}[]{0.45\textwidth}
                \epsscale{1.0}
                \plotone{fd1f.eps}
        \end{minipage}
        \begin{minipage}[]{0.45\textwidth}
                \epsscale{1.0}
                \plotone{fd1g.eps}
        \end{minipage}
        \begin{minipage}[]{0.45\textwidth}
                \epsscale{1.0}
                \plotone{fd1h.eps}
        \end{minipage}
\end{center}
\caption
{The diversity patterns from the companion-induced asymmetric configuration. Shown here is the residual of the synthetic spectra for Model RGa after being divided by the mean spectrum, shown for the optical range (left) and for the NIR (right). The epoch is 16.8 days after the explosion. The panels are divided into four according to the viewing direction ($\theta = 0$ to $\theta = \pi$, from bottom to top). The synthetic spectra are binned within 3 time bins, but no additional binning is performed in the wavelength and viewing angle directions. 
\label{figd1}}
\end{figure*}

\clearpage
\begin{figure*}
\begin{center}
        \begin{minipage}[]{0.45\textwidth}
                \epsscale{1.0}
                \plotone{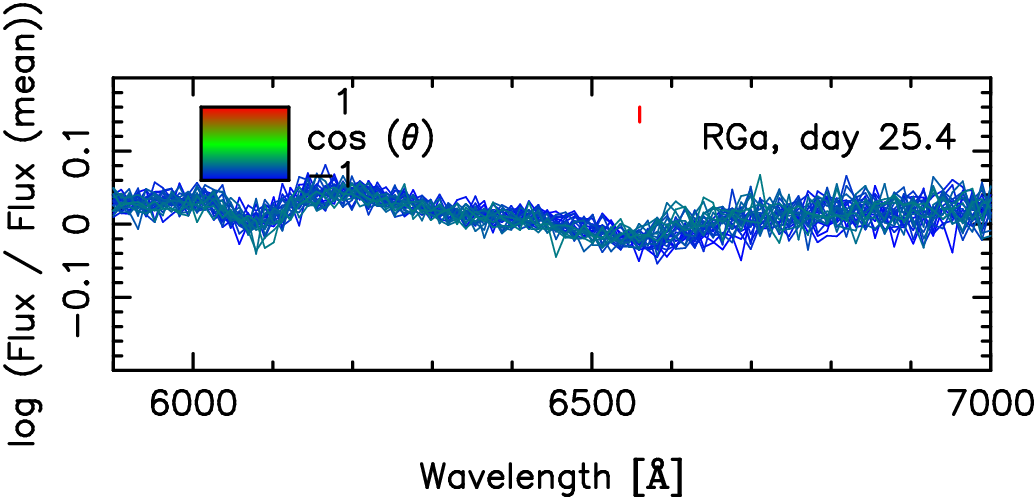}
        \end{minipage}
        \begin{minipage}[]{0.45\textwidth}
                \epsscale{1.0}
                \plotone{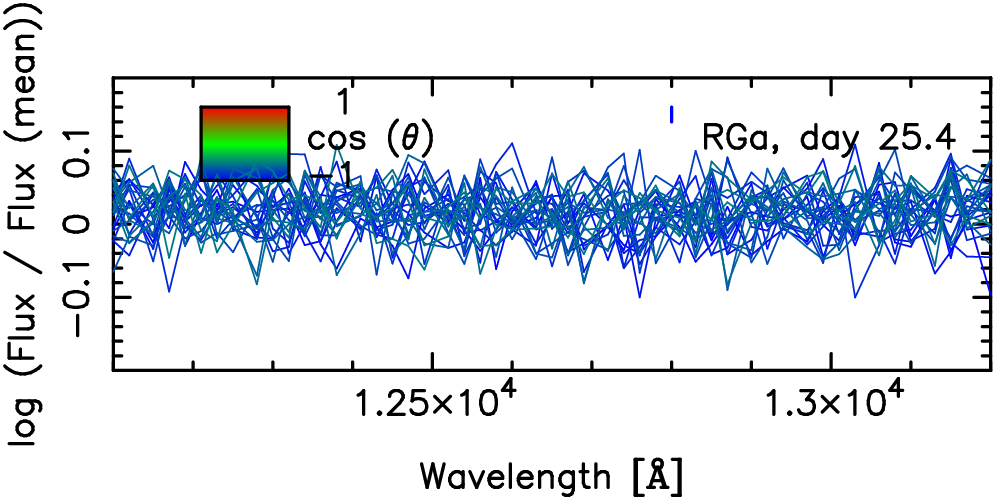}
        \end{minipage}
        \begin{minipage}[]{0.45\textwidth}
                \epsscale{1.0}
                \plotone{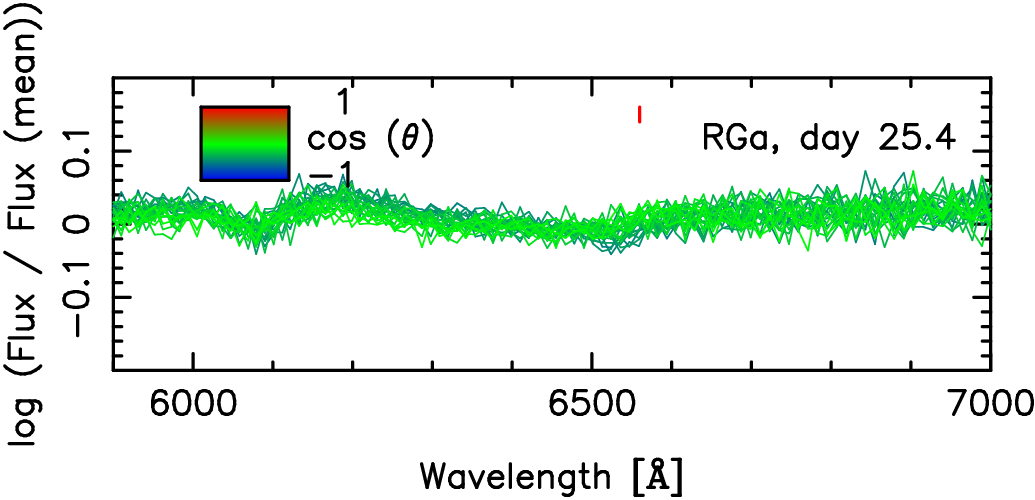}
        \end{minipage}
        \begin{minipage}[]{0.45\textwidth}
                \epsscale{1.0}
                \plotone{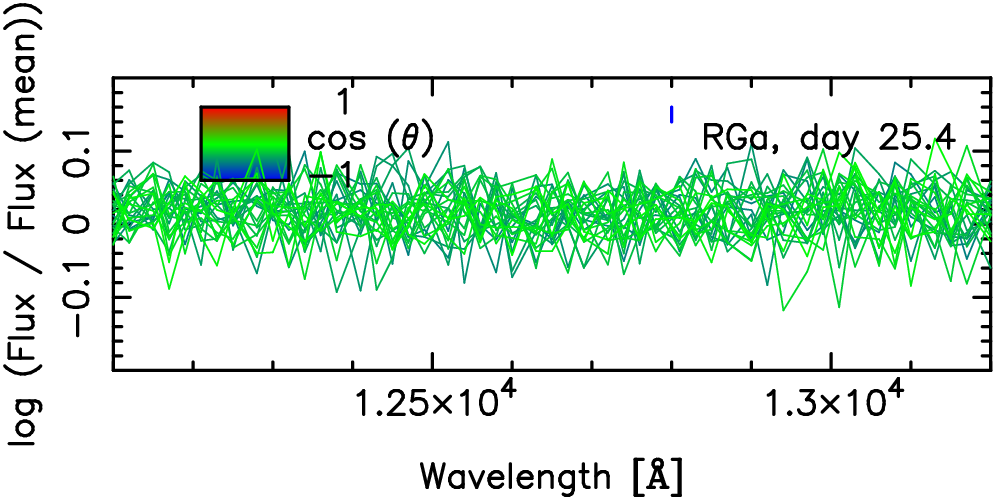}
        \end{minipage}
        \begin{minipage}[]{0.45\textwidth}
                \epsscale{1.0}
                \plotone{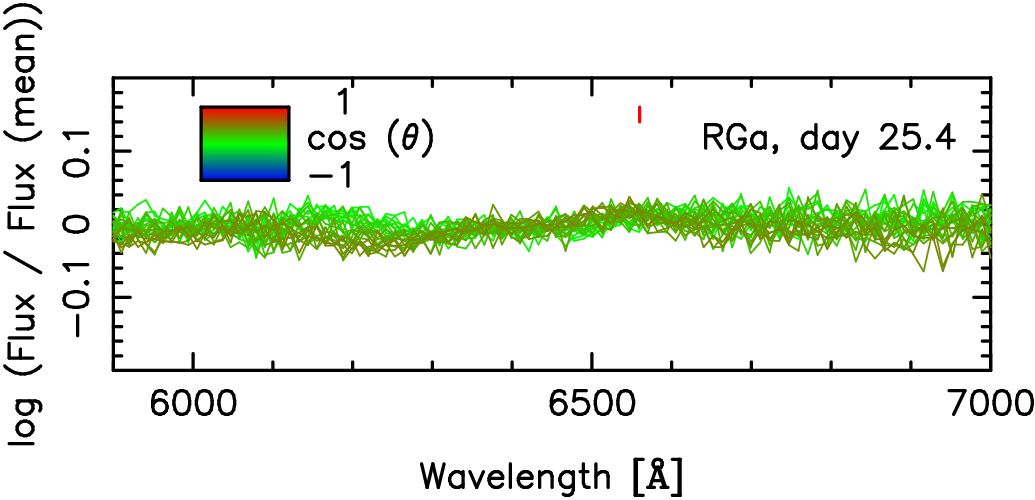}
        \end{minipage}
        \begin{minipage}[]{0.45\textwidth}
                \epsscale{1.0}
                \plotone{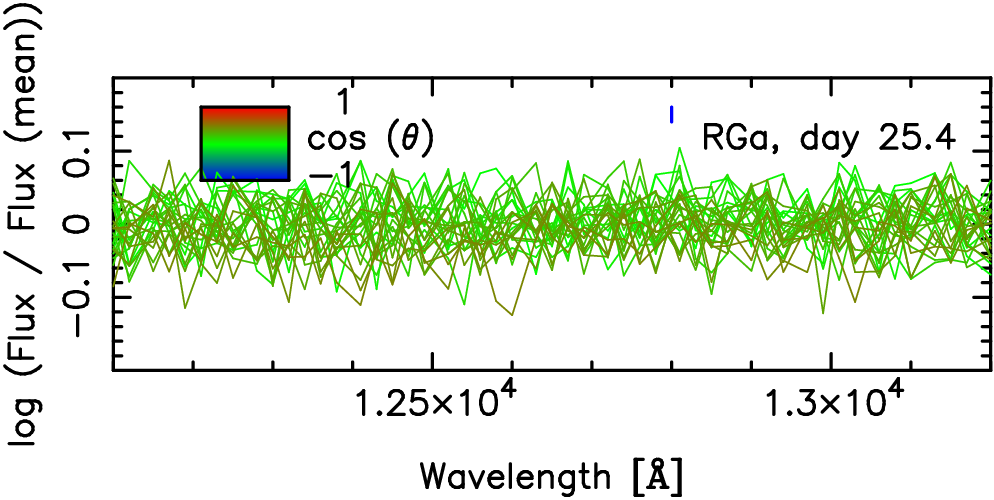}
        \end{minipage}
        \begin{minipage}[]{0.45\textwidth}
                \epsscale{1.0}
                \plotone{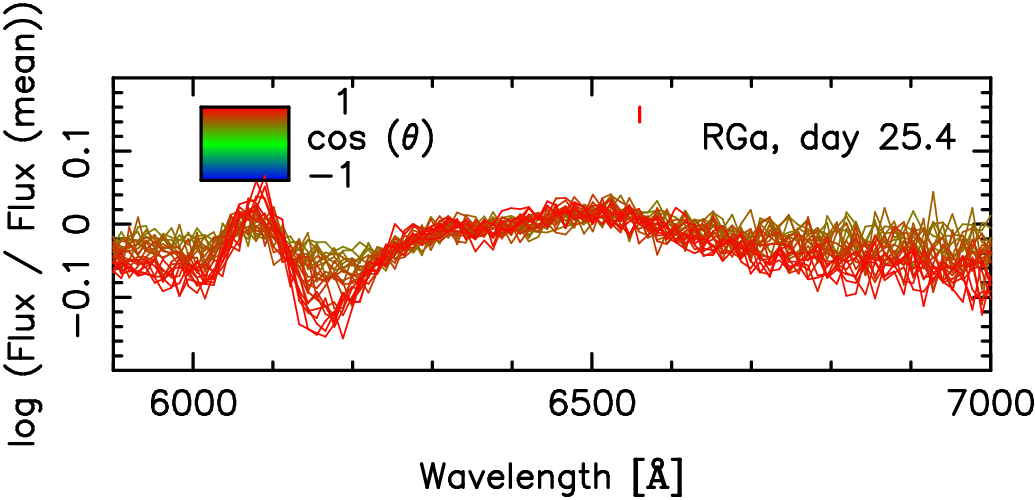}
        \end{minipage}
        \begin{minipage}[]{0.45\textwidth}
                \epsscale{1.0}
                \plotone{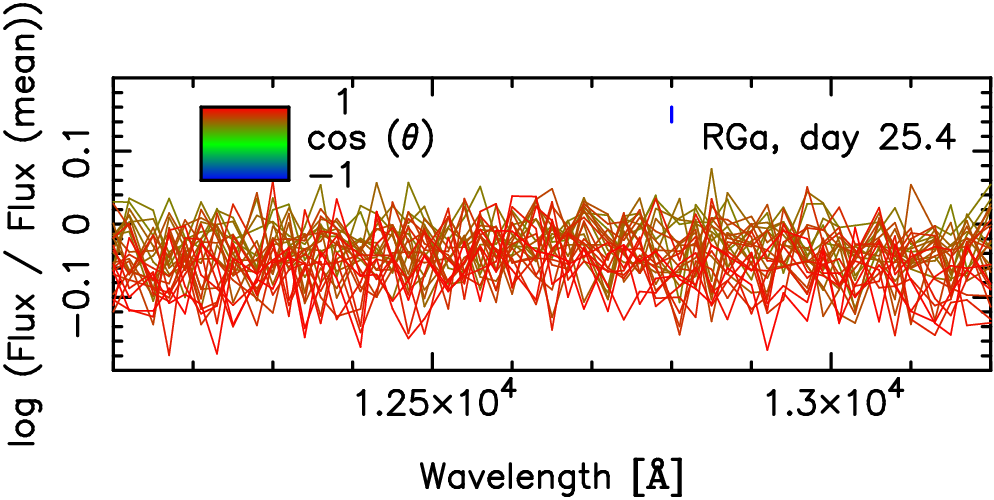}
        \end{minipage}
\end{center}
\caption
{The diversity patterns from the companion-induced asymmetric configuration, as is the same with Figure D1. The epoch is 25.4 days since the explosion. 
\label{figd2}}
\end{figure*}

\clearpage
\begin{figure*}
\begin{center}
        \begin{minipage}[]{0.45\textwidth}
                \epsscale{1.0}
                \plotone{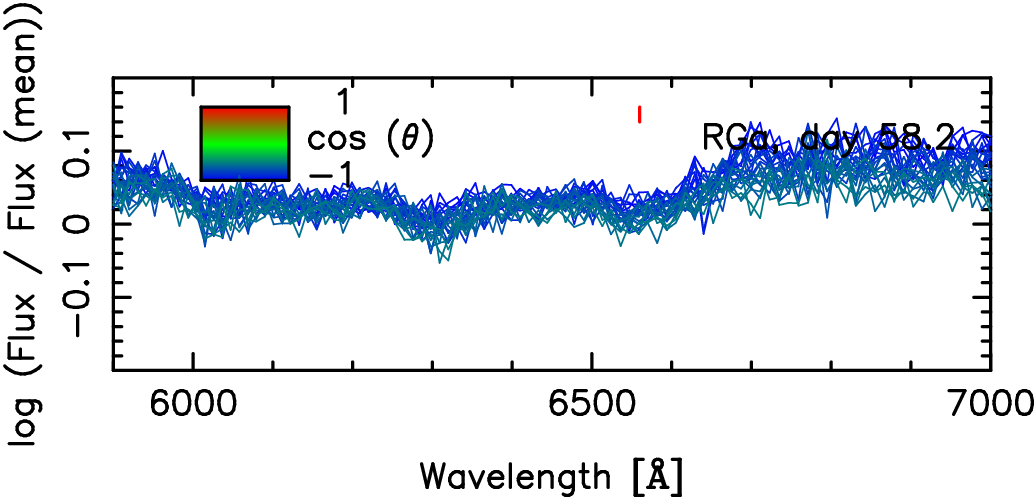}
        \end{minipage}
        \begin{minipage}[]{0.45\textwidth}
                \epsscale{1.0}
                \plotone{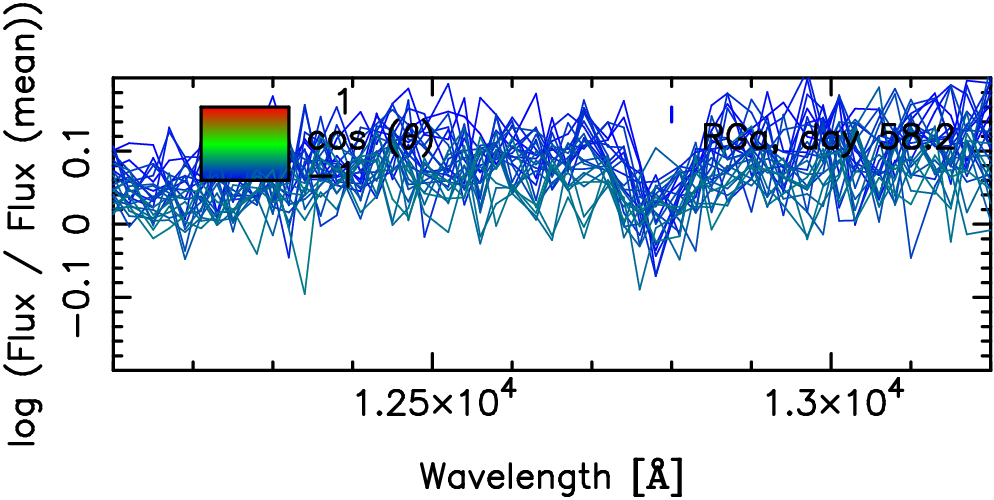}
        \end{minipage}
        \begin{minipage}[]{0.45\textwidth}
                \epsscale{1.0}
                \plotone{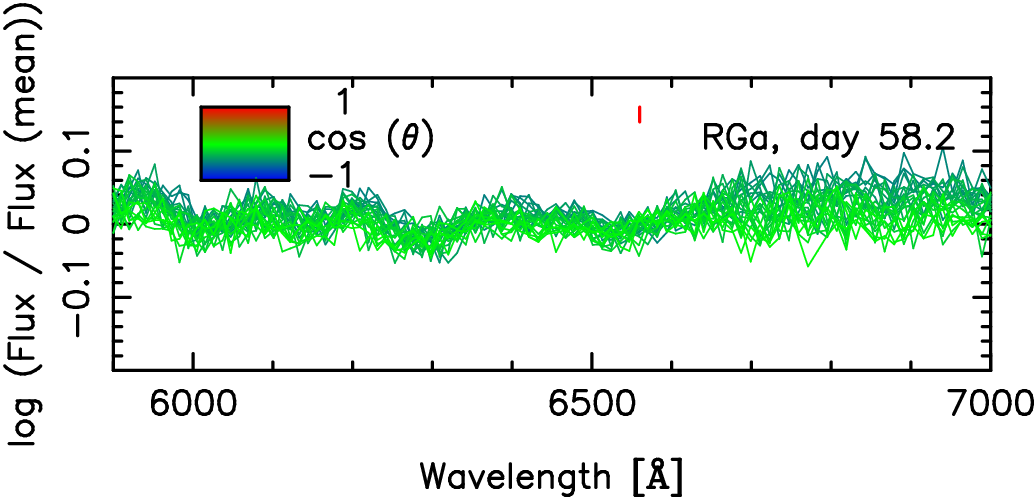}
        \end{minipage}
        \begin{minipage}[]{0.45\textwidth}
                \epsscale{1.0}
                \plotone{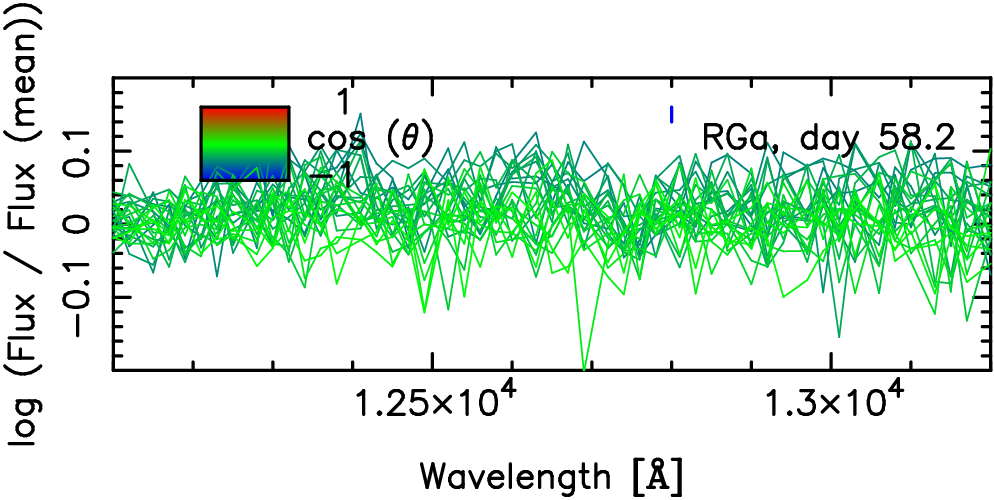}
        \end{minipage}
        \begin{minipage}[]{0.45\textwidth}
                \epsscale{1.0}
                \plotone{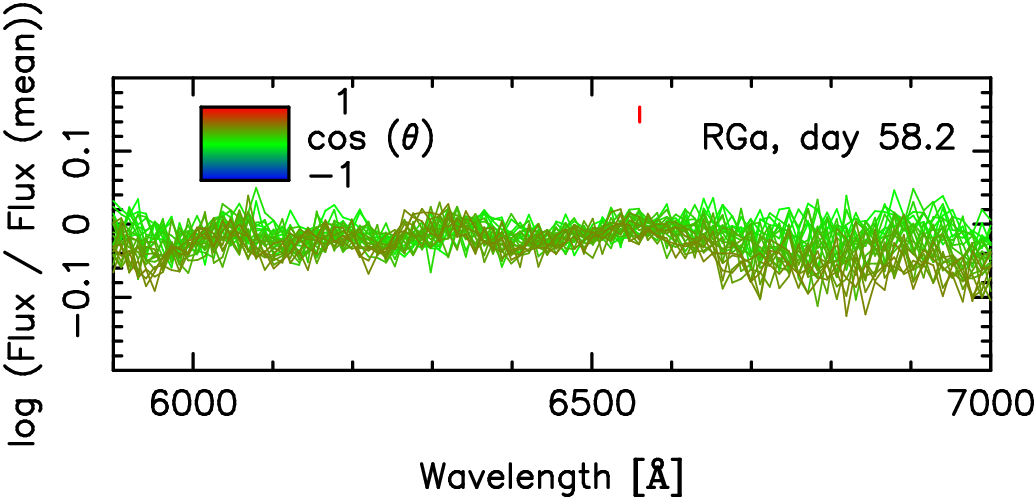}
        \end{minipage}
        \begin{minipage}[]{0.45\textwidth}
                \epsscale{1.0}
                \plotone{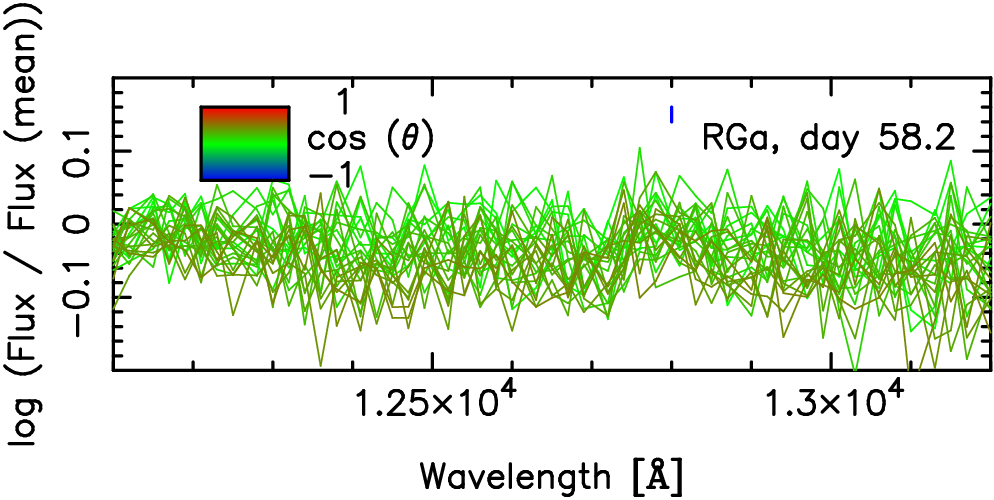}
        \end{minipage}
        \begin{minipage}[]{0.45\textwidth}
                \epsscale{1.0}
                \plotone{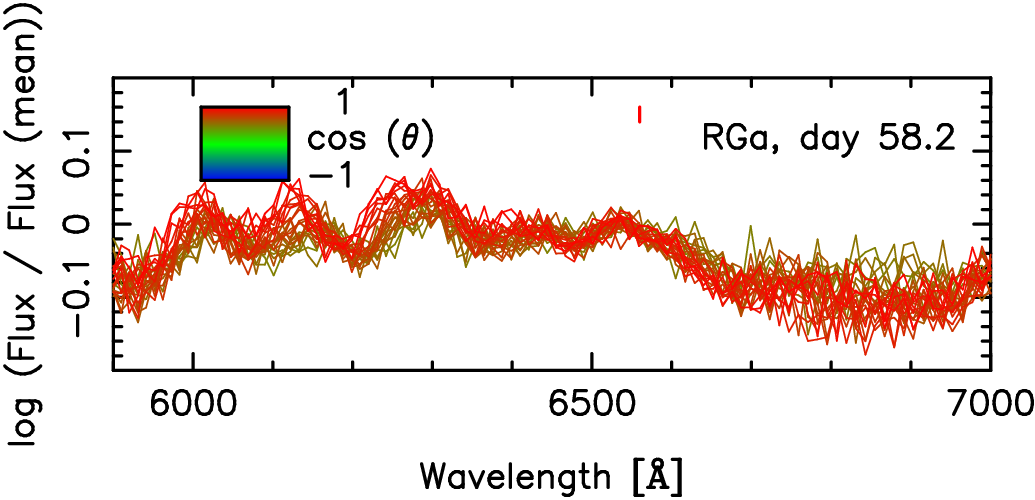}
        \end{minipage}
        \begin{minipage}[]{0.45\textwidth}
                \epsscale{1.0}
                \plotone{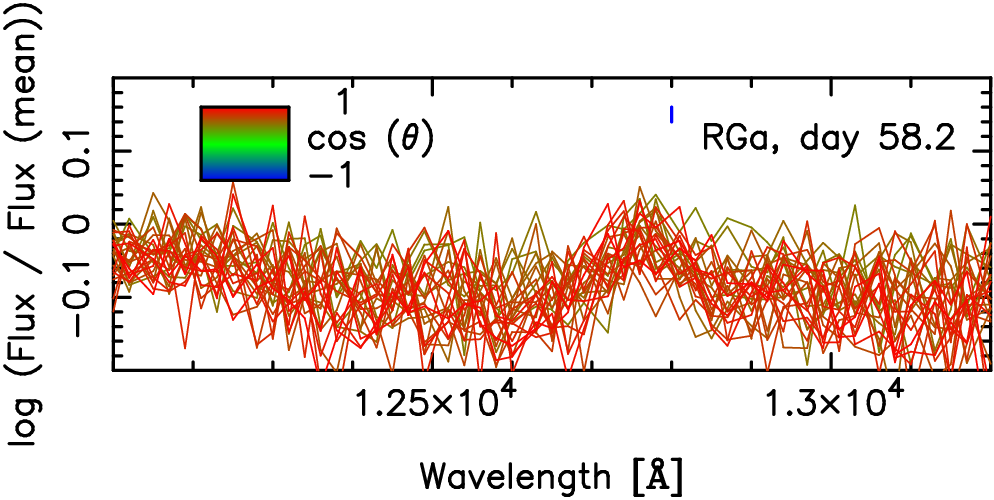}
        \end{minipage}
\end{center}
\caption
{The diversity patterns from the companion-induced asymmetric configuration, as is the same with Figure D1. The epoch is 58.2 days since the explosion. 
\label{figd3}}
\end{figure*}

\end{document}